\documentclass{emulateapj}
\usepackage{lscape}
\newcommand{\degree}{\hbox{$^\circ$}}
\newcommand{\dgr}{$\cal{D}$}
\newcommand{\etal}{et\,al.}
\newcommand{\halpha}{H$\alpha$}
\newcommand{\gsim}{\raise0.3ex\hbox{$>$}\kern-0.75em{\lower0.65ex\hbox{$\sim$}}}
\newcommand{\iras}{{\it IRAS}}
\newcommand{\jypbeam}{\,\,Jy\,Bm$^{-1}$}
\newcommand{\kms}{km\,s$^{-1}$}
\newcommand{\lsim}{\raise0.3ex\hbox{$<$}\kern-0.75em{\lower0.65ex\hbox{$\sim$}}}
\newcommand{\lsun}{L$_{\odot}$}
\newcommand{\msun}{M$_{\odot}$}
\newcommand{\mm}{$\mu$m}
\newcommand{\pom}{\,$\pm$\,}
\newcommand{\sings}{{\it SINGS}}
\newcommand{\spitzer}{{\it Spitzer}}

\newcommand{\HI}{H~{\sc i}}
\newcommand{\HII}{H~{\sc ii}}
 
\newcommand{\zsun}{Z$_{\odot}$}
\begin{document}     
\slugcomment{Accepted for publication in the Astrophysical Journal}
\title{The Nature of Infrared Emission in the Local Group Dwarf Galaxy NGC 
6822 As Revealed by {\it Spitzer}}


\author{John M. Cannon\altaffilmark{1,2},
Fabian Walter\altaffilmark{2},
Lee Armus\altaffilmark{3},
George J. Bendo\altaffilmark{4,5},
Daniela Calzetti\altaffilmark{6},
Bruce T. Draine\altaffilmark{7},
Charles W. Engelbracht\altaffilmark{5},
George Helou\altaffilmark{3},
Robert C. Kennicutt, Jr.\altaffilmark{8,5},
Claus Leitherer\altaffilmark{6},
H{\'e}l{\`e}ne Roussel\altaffilmark{2}
Caroline Bot\altaffilmark{3},
Brent Buckalew\altaffilmark{3},
Daniel A. Dale\altaffilmark{9},
W.~J.~G. de Blok\altaffilmark{10},
Karl D. Gordon\altaffilmark{5},
David J. Hollenbach\altaffilmark{11},
Thomas H. Jarrett\altaffilmark{3},
Martin J. Meyer\altaffilmark{6},
Eric J. Murphy\altaffilmark{12}
Kartik Sheth\altaffilmark{3},
Michele D. Thornley\altaffilmark{13}
}

\altaffiltext{1}{Astronomy Department, Wesleyan University, 
Middletown, CT 06457; cannon@astro.wesleyan.edu}
\altaffiltext{2}{Max-Planck-Institut f{\"u}r Astronomie, K{\"o}nigstuhl 17,
D-69117, Heidelberg, Germany; walter@mpia.de, roussel@mpia.de}
\altaffiltext{3}{California Institute of Technology, MC 314-6, Pasadena, CA
91101; lee@ipac.caltech.edu, gxh@ipac.caltech.edu, bot@caltech.edu,
brentb@ipac.caltech.edu, jarrett@ipac.caltech.edu, kartik@astro.caltech.edu}
\altaffiltext{4}{Astrophysics Group, Imperial College, Blackett Laboratory,
Prince Consort Road, London SW7 2AZ United Kingdom; g.bendo@imperial.ac.uk}
\altaffiltext{5}{Steward Observatory, University of Arizona, 933 North Cherry
Avenue, Tucson, AZ 85721; chad@as.arizona.edu, kgordon@as.arizona.edu}
\altaffiltext{6}{Space Telescope Science Institute, 3700 San Martin Drive,
Baltimore, MD 21218; calzetti@stsci.edu, leitherer@stsci.edu, martinm@stsci.edu} 
\altaffiltext{7}{Princeton University Observatory, Peyton Hall, Princeton, NJ
08544; draine@astro.princeton.edu}
\altaffiltext{8}{Institute of Astronomy, University of Cambridge, Madingley 
Road, Cambridge CB3 0HA, UK; robk@ast.cam.ac.uk}
\altaffiltext{9}{Department of Physics and Astronomy, University of Wyoming,
Laramie, WY 82071; ddale@uwyo.edu}
\altaffiltext{10}{Research School of Astronomy \& Astrophysics, Mount Stromlo
Observatory, Cotter Road, Weston ACT 2611, Australia; edeblok@mso.anu.edu.au}
\altaffiltext{11}{NASA/Ames Research Center, MS 245-6, Moffett Field, CA,
94035; hollenba@ism.arc.nasa.gov}
\altaffiltext{12}{Department of Astronomy, Yale University, New Haven, CT 
06520; murphy@astro.yale.edu}
\altaffiltext{13}{Department of Physics and Astronomy, Bucknell University, 
Lewisburg, PA 17837; mthornle@bucknell.edu}

\begin{abstract}

We present {\it Spitzer} imaging of the metal-deficient
(Z\,$\simeq$30\%\,\zsun) Local Group dwarf galaxy NGC\,6822.  On
spatial scales of $\sim$130 pc, we study the nature of IR, \halpha,
\HI, and radio continuum emission.  Nebular emission strength
correlates with IR surface brightness; however, roughly half of the IR
emission is associated with diffuse regions not luminous at \halpha\
(as found in previous studies).  The global ratio of dust to \HI\ gas
in the ISM, while uncertain at the factor of $\sim$2 level, is $\sim$25
times lower than the global values derived for spiral galaxies using
similar modeling techniques; localized ratios of dust to \HI\ gas are
about a factor of five higher than the global value in NGC\,6822.
There are strong variations (factors of $\sim$ 10) in the relative
ratios of \halpha\ and IR flux throughout the central disk; the low
dust content of NGC\,6822 is likely responsible for the different
\halpha/IR ratios compared to those found in more metal-rich
environments.  The \halpha\ and IR emission is associated with
high-column density ($\gsim$10$^{21}$ cm$^{-2}$) neutral gas.
Increases in IR surface brightness appear to be affected by both
increased radiation field strength and increased local gas density.
Individual regions and the galaxy as a whole fall within the observed
scatter of recent high-resolution studies of the radio-far IR
correlation in nearby spiral galaxies; this is likely the result of depleted
radio and far-IR emission strengths in the ISM of this dwarf galaxy.

\end{abstract}						


\section{Introduction}
\label{S1}

Nearby dwarf galaxies provide a unique opportunity to resolve the interaction
of the multiple phases of the interstellar medium (ISM).  Star formation in
these systems is more susceptible to small-scale processes (e.g., turbulence,
feedback) than to the galaxy-wide perturbations (i.e., spiral density waves)
that are common in more massive galaxies.  With their typically low nebular
metallicities, star-forming dwarf galaxies serve as fiducial examples for
comparison with the high-redshift star-forming systems predicted in the cold
dark matter paradigm \citep[e.g.,][]{babul96,ellis97,mateo98}.

\begin{figure*}[!ht]
\plotone{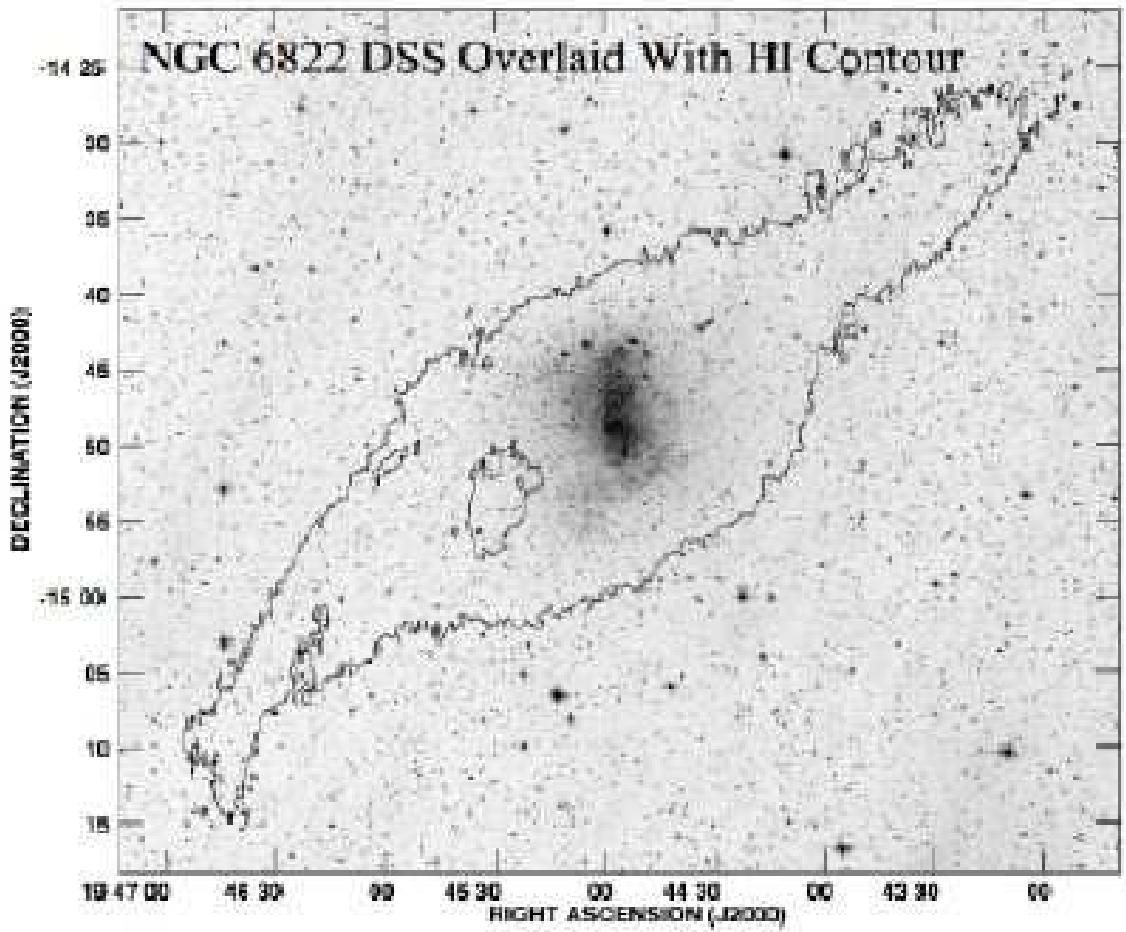}
\caption{DSS image of NGC\,6822, overlaid with a contour of \HI\ coulmn density at the 
       2\,$\times$\,10$^{20}$ cm$^{-2}$ level.  The \HI\ disk is much larger than the 
       high surface brightness optical body, which was the target of our IRAC and MIPS 
       observations.  However, \citet{deblok06} find an extended stellar population that
       exceeds the size of the \HI\ disk (see further discussion in \S~\ref{S1}).}
\label{figcap1}
\end{figure*}

The detailed processes of heating and cooling of the ISM are
fundamental to our understanding of the nature of galaxy evolution.
An environmentally-dependent fraction of UV photons produced by
massive stars will be absorbed by dust and gas and re-radiated in the
far-infrared (FIR) or in nebular emission lines.  This fraction will
depend on various factors, including the local dust-to-gas ratio
(related to the metallicity and hence the dust content) and the
porosity of the local ISM.  In cases where the dust and stars are well
mixed, and where the extinction is moderate, there will exist a
correlation between the fluxes in various wavelength bands, including
the ultraviolet (UV), \halpha, FIR, and radio continuum.  Nearby dwarf
galaxies offer a unique opportunity to study one of the regimes where
these correlations may begin to break down.

NGC\,6822 is a Local Group (D $=$ 490\,$\pm$\,40 kpc; {Mateo
1998}\nocite{mateo98}) dwarf irregular galaxy with a nebular abundance
Z $\simeq$ 30\% \zsun\ \citep{skillman89,lee06a}.  The total optical
luminosity is $\sim$9.4\,$\times$\,10$^7$ \lsun\ \citep{mateo98},
while the total \HI\ mass is 1.34\,$\times$\,10$^{8}$ \msun\ (see
detailed discussion in \S~\ref{S2.2} and {de~Blok \& Walter
2006}\nocite{deblok06}).  The system contains distributed low-level
star formation throughout the disk as well as actively star-forming
regions with associated \halpha\ emission which were first studied by
\citet{hubble25}.  The star formation has been quiescent over the past
few Gyr, without major peaks or dips in the absolute rate
\citep{wyder01,wyder03}.  \citet{deblok06} derive a total \halpha\ flux
of 2\,$\times$\,10$^{39}$ erg\,s$^{-1}$, corresponding to a 
global \halpha-based star formation rate
(SFR) of $\sim$ 0.015 \msun\,yr$^{-1}$ using the calibration of 
\citet{kennicutt98}.

\begin{figure*}[!ht]
\plotone{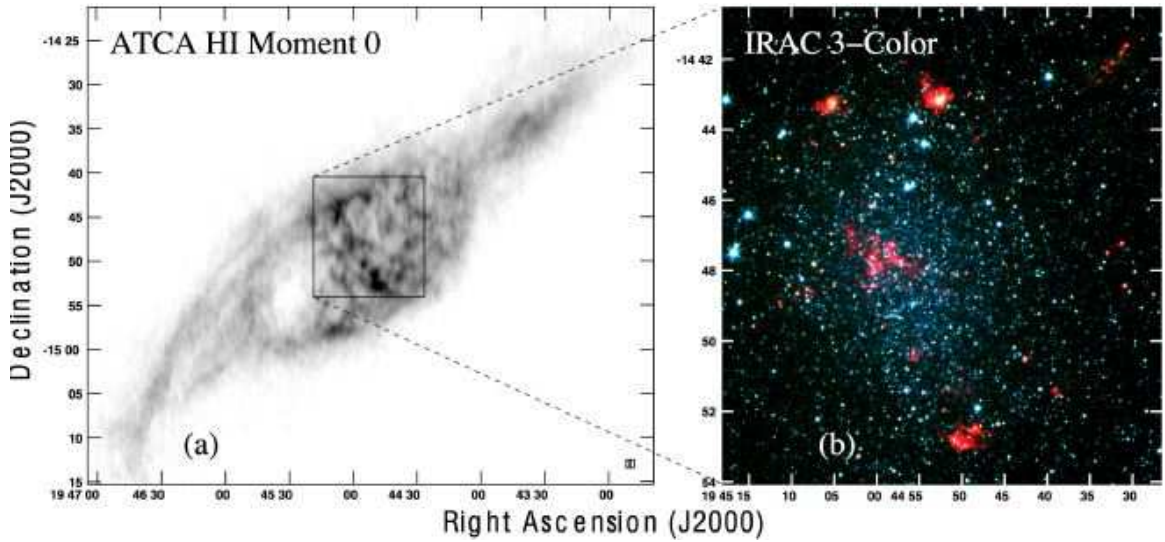}
\caption{\HI\ (a) and IRAC (b) images of NGC\,6822. (a) shows the
    total \HI\ column density distribution (de~Blok \& Walter 2000),
    while (b) shows the underlying stellar and warm dust
    components. The three-color image shows the 3.6 \mm\ band as blue,
    the 4.5 \mm\ band as green, and the 8 \mm\ band as red; regions of
    hot dust emission, indicative of active star formation, appear as
    diffuse regions of red emission. The box in (a) shows the
    approximate field of view shown in (b).}
\label{figcap2}
\end{figure*}

NGC\,6822 boasts an unusual \HI\ distribution; in the extended \HI\
disk (see Figure~\ref{figcap1}) lies one of the largest known holes in
the ISM of a dwarf galaxy, first discussed by \citet{deblok00}.
\citet{deblok06} argue that this hole is likely caused by the combined
effects of stellar evolution.  That study also reveals the presence of
a significant blue star population in the outer \HI\ disk, and a lower
surface brightness, spheroidal red stellar population extending beyond
the \HI\ distribution.  The extended \HI\ disk may include a dwarf
companion galaxy in the northwest region (note that interactions with
this system may have played a part in triggering some of the recent
star formation in the galaxy; see {de~Blok \& Walter
2000}\nocite{deblok00}).  The well-sampled rotation curve shows that
the system is highly dark-matter dominated \citep{weldrake03}.  The
proximity of NGC\,6822 means that it is one of only a few places where
the small-scale ($\sim$100 pc) structure of the ISM can be studied at
multiple wavelengths.

Previous infrared studies of NGC\,6822 have examined the heating and
cooling mechanisms in the ISM.  Using {\it InfraRed Astronomical
Satellite} ({\iras}) data, \citet{gallagher91} found a variable
\halpha/FIR ratio within the disk, with roughly 50\% of the FIR
emission arising from regions that are luminous at \halpha.  The
remaining fraction of diffuse IR flux was attributed to a low optical
depth (i.e., low dust content) in the ISM.  Thus, regions of strong IR
emission correspond to comparatively dusty regions that also contain a
younger stellar population.  A second \iras\ study by \citet{israel96}
found that three of the bright IR peaks were associated with luminous
\HII\ regions, while two equally bright sources were unassociated with
obvious sources of ionizing radiation.  These authors confirm the
prominent diffuse dust continuum emission found by
\citet{gallagher91}.  Further, they find, using single-temperature
blackbody fits to the far-IR data, a dust-to-gas ratio
(1.4\,$\times$\,10$^{-4}$) that is much lower than typical values seen
in more metal-rich systems such as the Milky Way (average dust-to-gas
ratios of $\sim$0.006--0.01; {Sodroski \etal\
[1997]}\nocite{sodroski97}, {Li [2004]}\nocite{li04}).  NGC\,6822 was
observed with the {\it Infrared Satellite Observatory} at 6.75 and 15
\mm, but only the most luminous \HII\ regions were detected
\citep{dale00,hunter01}.

In this paper, we present \spitzer\ Infrared Array Camera (IRAC) and
Multiband Imaging Photometer for \spitzer\ (MIPS) imaging
\citep[see][]{werner04,fazio04,rieke04} of the high optical surface
brightness component (i.e., the inner portion of the \HI\ disk) of
NGC\,6822 (see Figure~\ref{figcap2} for a comparison of the field of
view imaged and the size of the \HI\ disk).  These data were obtained
as part of the {\it Spitzer Infrared Nearby Galaxies Survey} (\sings;
see {Kennicutt \etal\ 2003}\nocite{kennicutt03}). As the nearest
galaxy in the \sings\ sample, these data probe the nature of mid- and
far-IR emission at unprecedented sensitivity and resolution ($\sim$130
pc; see further discussion below). We compare these \spitzer\ data to
sensitive \halpha, \HI\ and radio continuum imaging in order to study
the relation between \HI\ surface density and the locations and
intensities of star formation and IR emission.

\begin{figure*}[!ht]
\begin{center}
\plotone{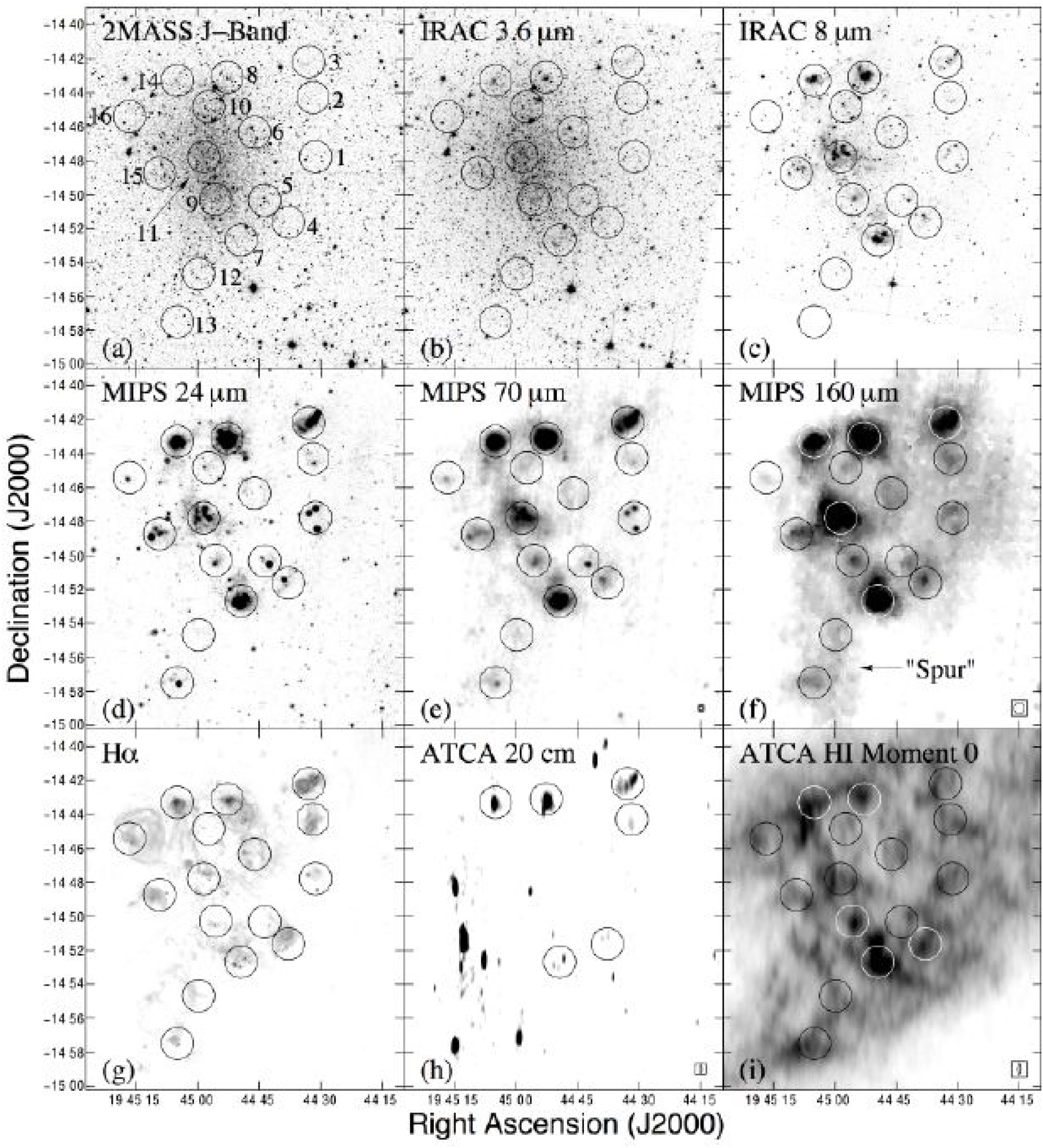}
\end{center}
\caption{2MASS J (a), 3.6 and 8 \mm\ (b, c), 24, 70, and 160 \mm\ (d,
   e, f), \halpha\ (g), 20\,cm radio continuum (h), and \HI\ spectral
   line (i) images of NGC\,6822.  All image scales are linear, except
   for the \halpha\ image (g), which is shown in a logarithmic stretch
   to highlight the low-surface brightness ionized gas that permeates
   much of the central disk of the system.  Beam sizes are shown as
   boxed ellipses for the MIPS and radio data, in the bottom right
   corners.  Overlaid are the sizes and locations of the smaller
   apertures [diameter $=$ 114\arcsec\ $\simeq$ 3\,$\times$\,FWHM (160
   \mm) $\simeq$ 270 pc] discussed in the text; these regions are labeled in panel
   (a).  Note that the regions that encompass the major IR emission
   complexes (regions 3, 7, 8, 11, and 14; see also Table~\ref{t1})
   have also been analyzed using apertures with four times the area.
   The 8 \mm\ image does not cover the full field of view shown in
   panel (c); since the 4.5 \mm\ band has the same orientation, flux
   densities are not available for region 13 in either of these
   bandpasses.  Note that the radio continuum image in panel (h) shows
   only the apertures where 20\,cm emission is detected; numerous
   background sources exist within this field of view (see further
   discussion in \S~\ref{S3.3}).  The aperture sizes are large enough
   to cover the bulk of the extended long-wavelength PSFs, minimizing
   aperture correction effects (see further discussion in
   \S~\ref{S3}).}
\label{figcap3}
\end{figure*}

\begin{deluxetable*}{lcccc}[ht]
\tablecaption{Infrared Sources in NGC\,6822}
\tablewidth{0pt}
\tablehead{
\colhead{Identification}
&\colhead{$\alpha$ (J2000)}
&\colhead{$\delta$ (J2000)}
&\colhead{Previous I.D.}
&\colhead{References}}
\startdata
NGC\,6822--1  &19:44:31.28 &$-$14:47:46.69 &I96\,1 &3\\
NGC\,6822--2  &19:44:31.92 &$-$14:44:16.70 &Hubble\,II &1\\
NGC\,6822--3  &19:44:32.96 &$-$14:42:10.71 &Hubble\,I/III, G91\,3, I96\,2 &1, 2\tablenotemark{a}, 3\\
NGC\,6822--4  &19:44:37.89 &$-$14:51:37.79 &I96\,3 &3\\
NGC\,6822--5  &19:44:43.69 &$-$14:50:19.86 &I96\,3 &3\\
NGC\,6822--6  &19:44:46.18 &$-$14:46:19.89 & &\\
NGC\,6822--7  &19:44:49.47 &$-$14:52:40.93 &Hubble\,IV, G91\,1, I96\,4 &1, 2, 3\\
NGC\,6822--8  &19:44:52.81 &$-$14:43:04.97 &Hubble\,V, G91\,3, I96\,5 &1, 2, 3\\
NGC\,6822--9  &19:44:55.69 &$-$14:50:17.00 & &\\
NGC\,6822--10 &19:44:57.36 &$-$14:44:50.01 & &\\
NGC\,6822--11 &19:44:58.59 &$-$14:47:47.02 &Hubble\,VI/VII, G91\,2, I96\,6 &1, 2, 3\\
NGC\,6822--12 &19:44:59.82 &$-$14:54:41.03 & &\\
NGC\,6822--13 &19:45:04.99 &$-$14:57:29.07 &I96\,7 &3\\
NGC\,6822--14 &19:45:05.01 &$-$14:43:17.07 &Hubble\,X, G91\,3, I96\,8 &1, 2, 3\\
NGC\,6822--15 &19:45:09.35 &$-$14:48:44.10 &Hubble\,IX, G91\,2, I96\,9 &1, 2, 3\\
NGC\,6822--16 &19:45:16.68 &$-$14:45:24.26 & &\\
\enddata
\label{t1}
\tablenotetext{a}{The sources in \citet{gallagher91} are not defined by
low-resolution (1.5\arcmin\,$\times$\,5\arcmin) \iras\ imaging; the
cross-identifications here correspond to only the highest-surface brightness
IR peaks in the IR or \halpha\ images.}  
\tablerefs{1 - \citet{hubble25}; 2 - \citet{gallagher91}; 3 - 
\citet{israel96}}
\end{deluxetable*}

\section{Observations and Data Reduction}
\label{S2}

\subsection{{\it Spitzer} Imaging}
\label{S2.1}

For an overview of the \sings\ observational strategies, see
\citet{kennicutt03}.  NGC\,6822 was observed for 122 minutes in IRAC
mosaicing mode on 2004, October 7 and 11; the separate visits to the
target ensure proper removal of image artifacts and asteroids.  The
observations cover the optical extent of the galaxy to the R$_{25}$
level (see Figures~\ref{figcap1} and \ref{figcap2}).  \citet{deblok06}
find that the stellar population extends throughout the entire \HI\
distribution; our IRAC data remain insensitive to stars in the outer
\HI\ disk of NGC\,6822.  The \sings\ IRAC pipeline processes the basic
calibrated data (BCD) images, producing mosaics with pixel scales of
0.75\arcsec\ and PSF FWHM values of 1.66\arcsec, 1.72\arcsec,
1.88\arcsec, and 1.98\arcsec\ at 3.6, 4.5, 5.8 and 8 \mm, respectively
\citep{fazio04}; note that 1\arcsec\ $=$ 2.4 pc at the adopted
distance of 490 kpc. Flux levels are uncertain at the $\sim$ 10\%
level, primarily due to systematic effects in the calibration process.

MIPS scan mapping mode observations were obtained on 2004, September
24 and 25, for a total of 162.1 minutes.  The MIPS Instrument Team
Data Analysis Tool \citep{gordon05} was used to process the BCD files,
producing mosaic images with pixel scales of 0.75\arcsec, 3.0\arcsec,
and 6.0\arcsec\ at 24, 70 and 160 \mm, respectively.  Systematic
uncertainties (e.g., detector nonlinearities, time-dependent
responsivity variations, background removal, etc.) limit the absolute
flux calibration to $\sim$ 10\% in the 24 \mm\ band and to $\sim$ 20\%
in the 70 and 160 \mm\ bands (note that the absolute calibration
changes slightly depending on which pipeline was used to produce the
BCD-level data\footnote{See the MIPS data analysis handbook for
details; http://ssc.spitzer.caltech.edu/mips}; the S10 calibration is
applied to these data). The FWHMs of the PSFs are 6\arcsec, 18\arcsec,
and 40\arcsec\ at 24, 70, and 160 \mm, respectively.  All flux
densities were extracted from the broadband images after convolution
to the 160 \mm\ beam.  We use convolution kernels that convert an input
PSF into a lower resolution output PSF using the ratio of Fourier
transforms of the output to input PSFs.  High frequency noise in the
input PSF is suppressed when these kernels are created.  See
K.~D. Gordon \etal\ (2006, in preparation) for details. 

At the Galactic latitude of NGC\,6822 ($-$14.8\degree), the foreground
stellar contamination is high.  The 3.6/8 \mm\ and 8/24 \mm\ colors
were examined in order to remove the brightest foreground sources,
which will have the most pronounced effect on derived flux densities.
Less luminous sources require a more detailed treatment; a first-order
correction can be estimated by computing the strength of foreground
star emission in regions that are clearly unassociated with the
stellar component of NGC\,6822, and scaling these to the sizes of the
regions under study \citep[e.g.,][]{lee06b}.  However, the proximity
of NGC\,6822 makes it large on the sky; the IRAC maps do not include a
large overscan region, making this technique susceptible to
small-number statistics.  We discuss the effects of foreground
contamination in more detail in \S~\ref{S3}.  Background sources are
more difficult to identify (though a first-order estimate of their
number density is available from some of the plots shown in this work;
see further discussion in \S~\ref{S3.2.1}), and no correction for
these sources has been explicitly applied.

Foreground Milky Way cirrus contamination is non-negligible in this
region of the sky \citep[e.g., ][]{gallagher91}; given the extent of
the galaxy, there will be variations in the cirrus contamination
levels within the system that we are unable to remove.  In an attempt
to subtract the smooth diffuse component, high-surface brightness
emission that is clearly associated with the galaxy (by comparison
with images at multiple wavelengths) was masked, leaving an estimate
of the (spatially variable) background cirrus emission. This
background information was smoothed by a 6\arcmin\ boxcar function and
then interpolated over the area of the galaxy.  This smoothed cirrus
component was then subtracted from the images at 24, 70 and 160 \mm.
Comparison of flux measurements before and after the cirrus removal
shows that this correction is of order $\sim$10\%.  The subtracted
components at 70 and 160 \mm\ are in rough agreement (in terms of both
morphology and surface brightness) with the lower-resolution
COBE/DIRBE maps presented by \citet{schlegel98}.  Given the systematic
uncertainties in the long-wavelength bands (see above) and the
potential contamination by foreground cirrus emission, and assuming a
random error component of up to $\sim$ 10\%, these corrections imply a
total error budget of $\sim$ 15\% at 24 \mm, and $\sim$ 25\% at 70 and
160 \mm.

\subsection{\HI\ Spectral Line and Radio Continuum Imaging}
\label{S2.2}

The \HI\ moment zero (representing integrated column density) image
presented here, first published in \citet{deblok00}, was obtained with
the {\it Australia Telescope Compact Array}\footnote{The Australia
Telescope is funded by the Commonwealth of Australia for operation as
a National Facility managed by the Commonwealth Scientific and
Industrial Research Organisation.} ({\it ATCA}) in 1999 and 2000,
using the 375, 750D, 1.5A, 6A and 6D configurations.  Fifteen 12-hour
synthesis observations were taken in mosaicing mode, with the spectral
line correlator providing 0.8 \kms\ channel separation.

\begin{deluxetable*}{lcccccccccc}
\tabletypesize{\tiny}
\tablecaption{Observed Quantities in NGC\,6822\tablenotemark{a}}
\tablewidth{0pt}
\tablehead{
\colhead{Region\tablenotemark{b}}
&\colhead{\halpha}
&\colhead{{3.6 \mm}\tablenotemark{c}}
&\colhead{{4.5 \mm}\tablenotemark{c}}
&\colhead{{5.8 \mm}\tablenotemark{c}}
&\colhead{{8.0 \mm}\tablenotemark{c}}
&\colhead{24 \mm}
&\colhead{70 \mm}
&\colhead{160 \mm}
&\colhead{20 cm}
&\colhead{\HI}\\
  &(erg\,s$^{-1}$\,cm$^{-2}$) &(mJy) &(mJy) &(mJy) &(mJy) &(mJy) &(mJy) &(mJy) &(mJy) &(Jy\,\kms)\\
}
\startdata
\multicolumn{11}{c}{Aperture radii $=$ 57\arcsec\ $=$ 3\,$\times$\,(160
  \mm\ PSF FWHM \gsim\ 130 pc)}\\\hline
1  &(9.6\pom1.9)E-13    &24\pom3  &16\pom2 &16\pom2  &28\pom4   &100\pom16  &560\pom140   &2100\pom530  &$\lsim$1.0\pom0.3                  &12.9\pom1.3\\
2  &(26\pom5)E-13       &17\pom2  &12\pom2 &6\pom0.6 &15\pom2   &20\pom3    &470\pom120   &2100\pom530  &3.3\pom0.4                         &12.2\pom1.2\\
3  &(110\pom20)E-13     &17\pom2  &14\pom2 &11\pom2  &27\pom4   &110\pom17  &1700\pom430  &3700\pom930  &12\pom1.2                          &12.0\pom1.2\\
4  &(24\pom5)E-13       &31\pom4  &21\pom3 &20\pom3  &26\pom3   &32\pom5    &630\pom160   &2100\pom530  &3.1\pom0.4                         &14.8\pom1.5\\
5  &(5.0\pom1)E-13      &43\pom5  &29\pom3 &26\pom3  &27\pom3   &81\pom13   &600\pom150   &1600\pom400  &$\lsim$1.0\pom0.3                  &11.0\pom1.1\\
6  &(14\pom3)E-13       &55\pom6  &38\pom4 &34\pom4  &33\pom4   &17\pom3    &520\pom130   &2000\pom500  &$\lsim$1.0\pom0.3                  &10.1\pom1.0\\
7  &(24\pom5)E-13       &42\pom5  &29\pom3 &49\pom6  &84\pom10  &260\pom40  &2300\pom580  &4900\pom1300 &6.4\pom0.7                         &21.8\pom2.2\\
8  &(120\pom25)E-13     &49\pom6  &39\pom4 &65\pom8  &130\pom14 &780\pom120 &4300\pom1100 &6800\pom1700 &21\pom2.1                          &14.0\pom1.4\\
9  &(4.6\pom0.9)E-13    &74\pom8  &49\pom4 &47\pom5  &41\pom4   &23\pom4    &900\pom230   &2100\pom530  &$\lsim$1.0\pom0.3                  &13.7\pom1.4\\
10 &(4.8\pom1.0)E-13    &62\pom7  &43\pom4 &44\pom5  &41\pom5   &31\pom5    &900\pom230   &1700\pom430  &$\lsim$1.0\pom0.3                  & 9.1\pom0.9\\
11 &(13\pom3)E-13       &92\pom10 &60\pom7 &76\pom9  &100\pom12 &97\pom15   &2000\pom500  &5200\pom1300 &$\lsim$1.0\pom0.3                  &15.1\pom1.5\\
12 &(5.9\pom1.2)E-13    &30\pom3  &22\pom3 &18\pom2  &11\pom2   &8.7\pom1.3 &380\pom100   &1400\pom350  &$\lsim$1.0\pom0.3                  &11.0\pom1.1\\
13 &(6.8\pom1.4)E-13    &12\pom2  &N/A     &10\pom1  &N/A       &30\pom5    &480\pom120   &1700\pom430  &$\lsim$1.0\pom0.3                  &14.1\pom1.4\\
14 &(110\pom20)E-13     &35\pom4  &28\pom3 &32\pom4  &49\pom6   &190\pom29  &2400\pom600  &3800\pom950  &11\pom1.1                          &16.3\pom1.6\\
15 &(17\pom4)E-13     &54\pom6  &39\pom4 &41\pom5  &39\pom4   &75\pom12   &880\pom220   &2100\pom530  &$\lsim$1.0\pom0.3                    &10.6\pom1.1\\
16\tablenotemark{d} &(17\pom4)E-13       &26\pom3  &20\pom2 &16\pom2  &10\pom6 &9.4\pom1.4 &330\pom85   &610\pom160   &$\lsim$1.0\pom0.3    &11.6\pom1.2\\
\hline
\multicolumn{11}{c}{Aperture radii $=$ 114\arcsec\ $=$ 6\,$\times$\,(160
  \mm\ PSF FWHM) \gsim\ 260 pc }\\
\hline
\multicolumn{11}{c}{Total Galaxy\tablenotemark{e}}\\
\hline
Total &(700\pom140)E-13 &2.26\pom0.23 &1.44\pom0.15 &1.90\pom0.19  &1.87\pom0.19  &2.51\pom0.50  &53.2\pom15  &136.2\pom40.0  &69.4\pom14\tablenotemark{f} &2266\pom227\\
\enddata
\label{t2}
\tablenotetext{a}{All values derived without aperture corrections (see 
discussion in \S~\ref{S2} and \S~\ref{S3}); the 24, 70 and 160 \mm\ images have 
had a first-order estimate of the Galactic cirrus component removed.  The IRAC 
flux densities may have sizeable aperture corrections (especially in the 5.8 and 
8.0 \mm\ bands, where the corrections will be of order $\sim$10\% for the 
57\arcsec\ radius apertures, and as large as $\sim$25\% for the global flux density.}
\tablenotetext{b}{See Table 1 for source coordinates and cross-identifications.}
\tablenotetext{c}{Only the brightest foreground sources have been removed in deriving these values.
Using a limited number of apertures which sample regions clearly unassociated with NGC\,6822,
we derive the following potential contamination levels from unsubtracted foreground sources:
IRAC 3.6 \mm\ $=$ 11.2\pom4.2 mJy; IRAC 4.5 \mm\ $=$ 12.0\pom5.1 mJy; IRAC 5.8, 8.0 \mm\ have average foreground values consistent with zero).}
\tablenotetext{d}{Uncertain local background value in the IR.}
\tablenotetext{e}{\spitzer\ global flux densities listed in units of Jy.}
\tablenotetext{f}{Calculated as the sum of individual sources, using the larger aperture when available.}
\end{deluxetable*}

To assure accurate total column density calibration, the
interferometric data were combined with single-dish observations taken
with the 64-m Parkes telescope.  These data use the same channel
separation as the interferometric observations.  The
``zero-spacing''correction was applied in the MIRIAD\footnote{See
http://www.atnf.csiro.au/computing/software/miriad} environment.
Conditional transfer functions (i.e., blanking) were applied to the
final cube to create the integrated column density image presented
here.  The image has a beam size of 42.4\arcsec\,$\times$\,12\arcsec\
(1 \jypbeam\ $=$ 1190.5 K).  The final cube that was imaged to produce
the moment zero map has a 5\,$\sigma$ column density sensitivity of
1.6\,$\times$\,10$^{19}$ cm$^{-2}$; the peak \HI\ column density is
$\sim$ 3\,$\times$\,10$^{21}$ cm$^{-2}$, and the total \HI\ mass is
1.34\,$\times$\,10$^{8}$ \msun\ at the adopted distance.  More
detailed discussions of the \HI\ data, analysis and interpretation can
be found in \citet{deblok00}, \citet{weldrake03}, and
\citet{deblok06}.

During the acquisition of the {\it ATCA} \HI\ data, a second frequency
band was simultaneously observed; centered at 1.38 GHz, this 128 MHz
bandwidth channel provides a sensitive L-band radio continuum map of
NGC\,6822.  Note that these data have no zero-spacing correction
applied; the presence of comparatively strong background sources and
the weak nature of the intrinsic radio continuum in NGC\,6822 make
integrated total flux densities difficult to determine.  Single-dish
observations at other frequencies \citep{klein83,klein86} suggest an
integrated L-band flux density of 100\,$\pm$\,25 mJy \citep{israel96}.
Comparing with our interferometric flux density measurements (see
\S~\ref{S3} below) suggests that the high surface brightness radio
components account for $\sim$70\% of the (uncertain) total L-band flux
density.  The remainder may exist as a diffuse synchrotron component
(see more detailed discussion on the thermal/nonthermal decomposition
within the galaxy in \S~\ref{S3.3}).  The radio continuum data use the
same basic reductions as for the spectral line data, though the final
image has a smaller beam size of 30\arcsec\,$\times$\,8\arcsec; the
rms noise is $\sim$ 90 $\mu$Jy\,Bm$^{-1}$.  Note that the 20\,cm image
presented in Figure~\ref{figcap3}(h) shows only the apertures with
radio detections.  Further, there are numerous background radio
sources in the field (see also the discussion in \S~\ref{S3.3}).

\subsection{Deep \halpha\ Imaging}
\label{S2.3}

The \halpha\ image presented here was obtained with the 2.5m Isaac
Newton Telescope, using three pointings with the Wide Field Camera to
cover the entire \HI\ disk of NGC\,6822.  The integration time was 80
minutes per pointing; wide-filter R-band imaging was used to remove
the continuum.  Flux calibration was obtained by comparison with
published fluxes for high-surface brightness regions in
\citet{hodge88,hodge89}, after correction for foreground
reddening \citep[E(B$-$V) $=$ 0.24;][]{gallart96a}. The limiting flux
level is $\sim$ 3.7\,$\times$\,10$^{-18}$ erg\,s$^{-1}$\,cm$^{-2}$,
roughly a factor of ten deeper than the images presented by Hodge and
collaborators.  For a detailed discussion of the \halpha\ image
handling, see \citet{deblok06}.

\begin{figure*}[!ht]
\plotone{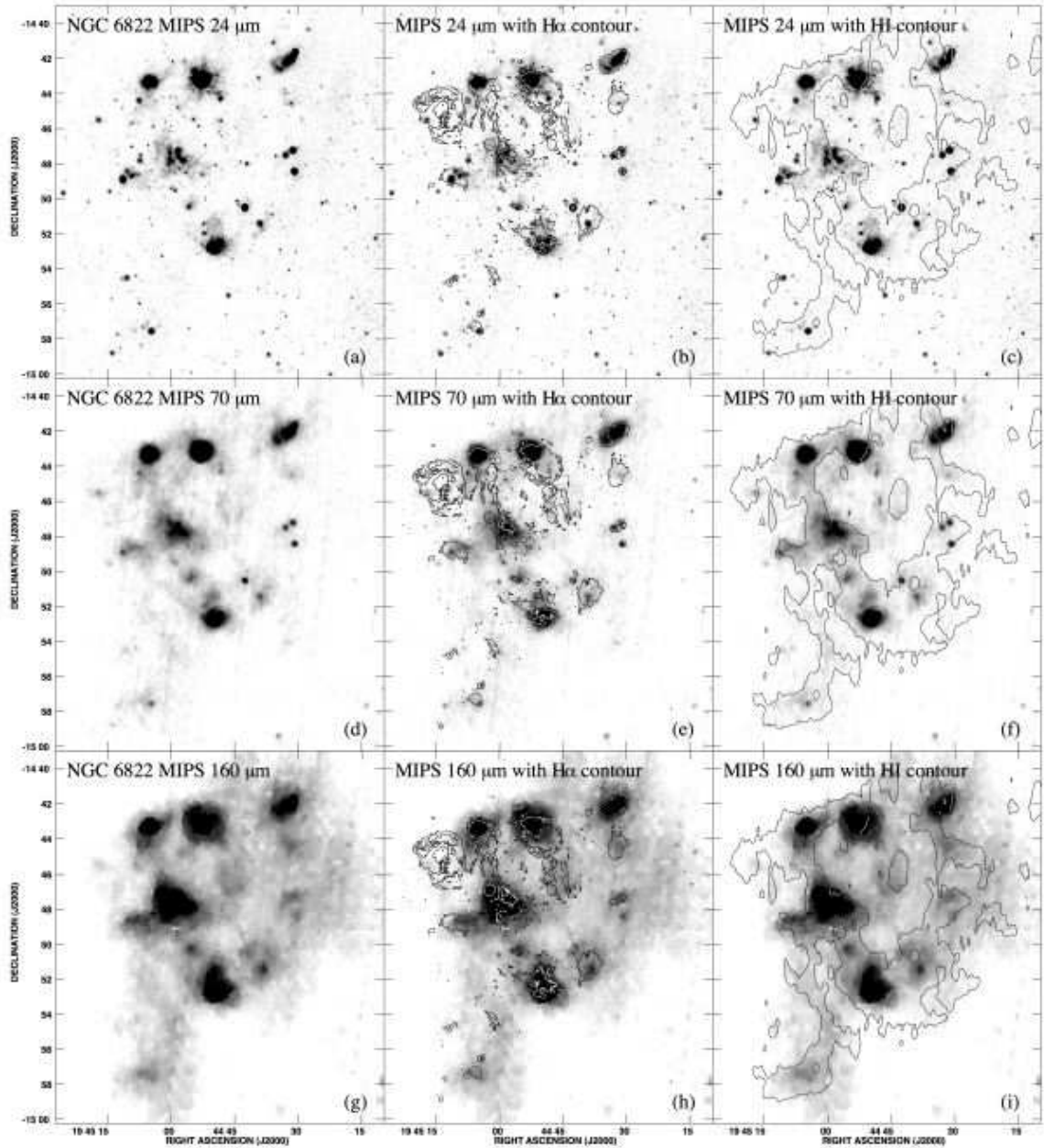}
\caption{The 24 (a, b, c), 70 (d, e, f) and 160 \mm\ images (g, h, i)
  of NGC\,6822, with \halpha\ (b, e, h) and \HI\ (c, f, i) contours
  overlaid.  The \halpha\ contour is at the flux level of
  3.8\,$\times$\,10$^{-18}$ erg\,s$^{-1}$\,cm$^{-2}$, highlighting
  low-level nebular emission throughout the disk; the \HI\ contour is
  at the level of 10$^{21}$ cm$^{-2}$, highlighting the canonical
  surface density threshold for star formation.  Note the striking
  correspondence between nebular, dust continuum, and \HI\
  emission.}
\label{figcap4}
\end{figure*}

\section{Multiwavelength Emission in NGC\,6822}
\label{S3}

Multiwavelength imaging of NGC\,6822 reveals a wealth of detail and
structure in the stellar and gaseous components.  In
Figure~\ref{figcap3} we present images of the central region of the
galaxy at nine different wavelengths.  In the 2MASS J
\citep{jarrett03} and 3.6 \mm\ bands [see Figures~\ref{figcap3}(a) and
\ref{figcap3}(b)], the red stellar component of the galaxy is smoothly
distributed, with few major stellar concentrations or clusters
(compare to {Wyder 2001, 2003}\nocite{wyder01,wyder03} and {de~Blok \&
Walter 2006}\nocite{deblok06} for more detailed views of the system at
optical wavelengths). Longward of 4 \mm, the spectral energy
distribution (SED) of the stellar population has dropped off
sufficiently that the warm dust emission from active star formation
regions dominates the galaxy morphology (see the 8 \mm\ image in
Figure~\ref{figcap3}(c).

\begin{figure*}[!ht]
\plotone{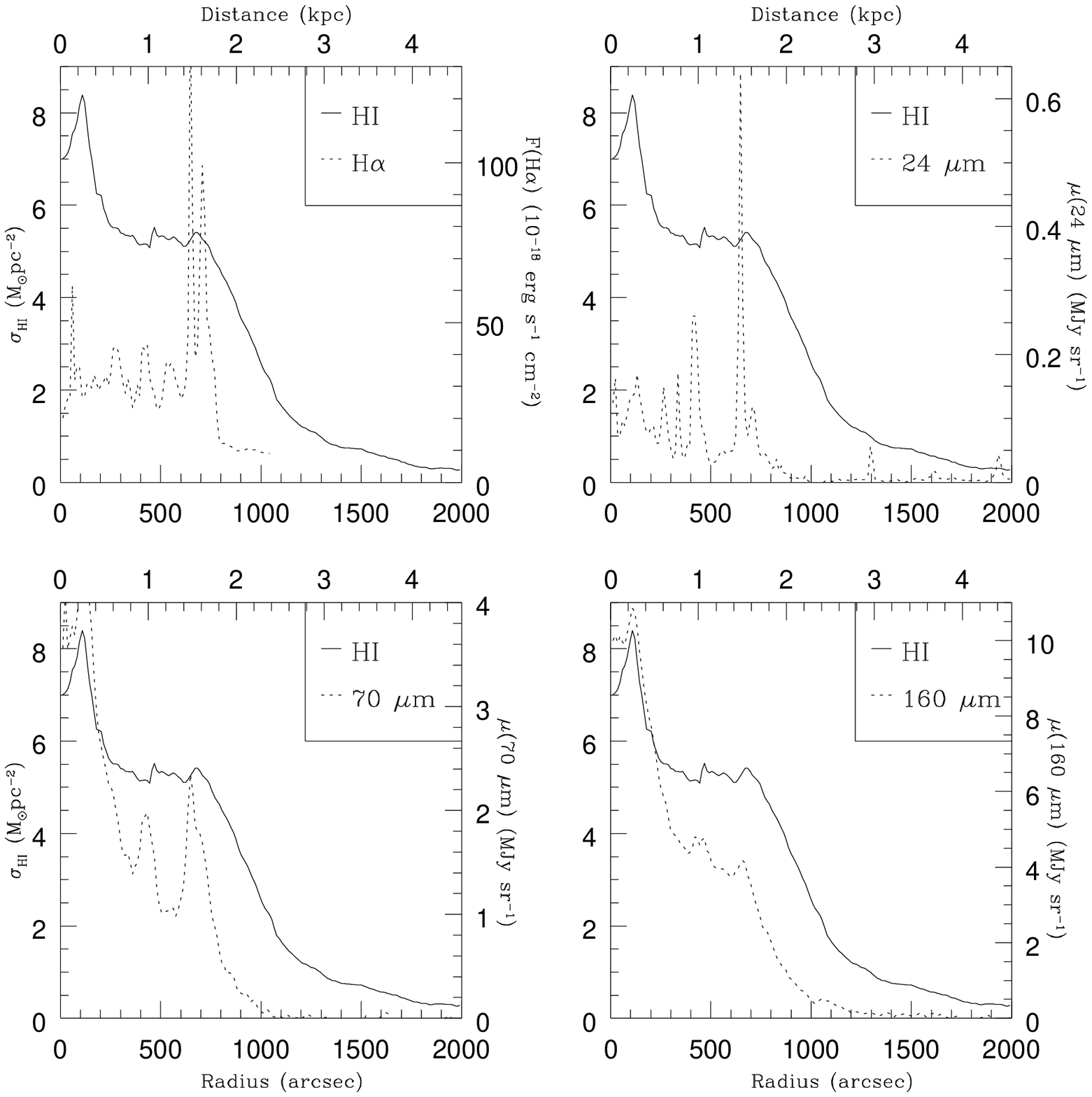}
\caption{Surface brightness profiles of \HI\ emission, overlaid with
   profiles of \halpha\ emission (a), and 24, 70 and 160 \mm\ emission
   (b, c, d).  The solid line and left-hand y-axis label in each panel
   correspond to the \HI\ surface density in units of \msun\,pc$^{-2}$
   (not inclination corrected); the dotted lines and right-hand y-axis
   labels correspond to the flux levels in \halpha\ (in units of
   erg\,sec$^{-1}$\,cm$^{-2}$) or to IR surface brightness values (in
   units of MJy\,sr$^{-1}$).  These profiles were created using the
   same parameters as the rotation curve analysis of
   \citet{weldrake03}; the central location is ($\alpha$, $\delta$)
   $=$ 19:44:58.04, $-$14:49:18.9 (J2000; this location is between
   apertures 9 and 11 in Figure~\ref{figcap3}), and an average
   position angle $\simeq$ 120\degree\ is used.  The profiles integrate 
   across the high-S/N ($\ge$ 3\,$\sigma$) regions of the images shown 
   in Figures~\ref{figcap3} and \ref{figcap4}.}
\label{figcap5}
\end{figure*}

Moving into the FIR, Figures~\ref{figcap3}d, e, and f show that the
active regions of star formation stimulate warm dust emission
throughout the system.  There is also very cool dust extending to the
south of the major star formation regions (though still well within
the optical and \HI\ radii; see Figure~\ref{figcap3}f); comparing this
emission with the \halpha\ and \HI\ images shown in
Figures~\ref{figcap3}(g) and \ref{figcap3}(i) leaves little doubt that
it is associated with NGC\,6822 and is not an image artifact.  Indeed,
both \citet{gallagher91} and \citet{israel96} recover this structure
in \iras\ maps. Further, \citet{deblok03} find that this region
(spatially coincident with the rim of the \HI\ hole) has the largest
concentration of blue stars in NGC\,6822.

The \halpha\ image presented in Figure~\ref{figcap3}(g) shows that
there is ionized gas throughout much of the central disk of the
galaxy; previous observations and comparisons with IR imaging have
concentrated on the higher-surface brightness star formation regions
\citep[e.g.,][]{gallagher91,israel96}.  Six of these regions
correspond to detections in the ATCA radio continuum image shown in
Figure~\ref{figcap3}(h).  Note that there are numerous background
sources in the radio continuum field of view; we have classified radio
``detections'' as areas of $>$ 5\,$\sigma$ flux density, that have
spatial counterparts in \halpha\ and the 24, 70 and 160 \mm\ bands
(note that we only show the apertures with radio detections in
Figure~\ref{figcap3}h).  Finally, the \HI\ distribution
(Figure~\ref{figcap3}i) throughout the central region of the galaxy
contains mostly high-surface brightness (i.e., \HI\ columns $\gsim$
10$^{21}$ cm$^{-2}$) emission.  There is clumping of the neutral gas
surrounding various emission peaks in other wavebands.

\begin{figure*}[!ht]
\plotone{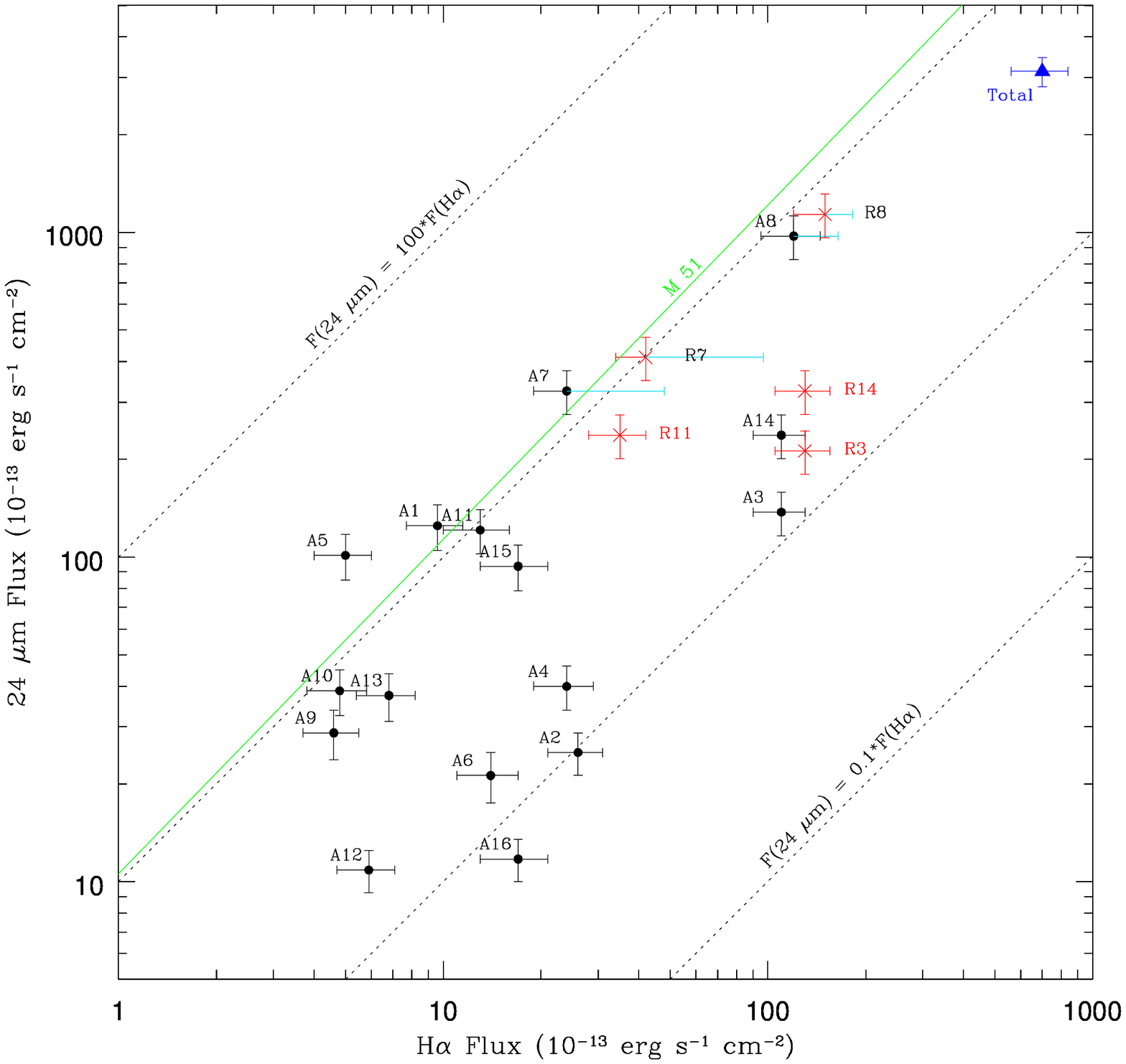}
\caption{Comparison of measured fluxes at \halpha\ and 24 \mm; black
   points indicate the 16 regions shown in Figure~\ref{figcap3}; red
   points show the five larger apertures used for the major IR peaks
   (see Table~\ref{t2}); the blue point shows photometry for the
   entire galaxy.  Asymmetric errorbars on points for regions 7 and 8
   (shown in cyan) demonstrate the magnitude of the measured
   extinctions at \halpha, as inferred from comparing (thermal) radio
   continuum and \halpha\ emission strengths (see discussion in
   \S~\ref{S3.1.1}).  The 24 \mm\ flux is calculated using the
   techniques discussed in {Calzetti \etal\ (2005)}, where F$_{\rm 24}
   =$ S$_{\rm 24}\times\nu_{\rm24}$; values of S$_{\rm 24}$ (in Jy)
   are taken from Table~\ref{t2}, and the frequency is evaluated at
   the wavelength of 24 \mm.  The relation derived by Calzetti \etal\
   (2005) for optically thick \HII\ regions in M\,51 is shown by the solid
   green line. The dotted lines show relations of direct
   proportionality, as labeled, for reference.}
\label{figcap6}
\end{figure*}

We study the panchromatic emission from 16 regions throughout the
galaxy; these regions are shown as circles overlaid on the various
panels of Figure~\ref{figcap3}, and their positions and
cross-identifications are given in Table~\ref{t1}.  The aperture
locations were chosen primarily to probe the IR emission complexes.
The aperture sizes and locations represent a compromise between
spatial resolution, minimization of aperture effects, and diversity of
regions that can be probed.  The aperture diameter of 114\arcsec\ ($=$
270 pc at the adopted distance) is $\sim$ 3\,$\times$ larger than the
FWHM of the 160 \mm\ PSF (and considerably larger than the FWHM of the
70 \mm, 24 \mm, and IRAC bands), which should alleviate serious
aperture corrections to the measured flux densities (especially at
3.6, 4.5, 24 and 70 \mm; extended-source aperture corrections at 5.8
and 8.0 are still uncertain, but may be as large as $\sim$30\%).  For
example, an aperture-centered point source at 160 \mm\ would have an
aperture correction of $\sim$ 30\%; this factor will be smaller for
extended emission which fills the beam (as is the case for most of the
160 \mm\ sources in the galaxy; however, the size of the correction
will depend on the IR surface brightness in the regions surrounding
the apertures as well).  This chosen aperture size is well-matched to
most of the discrete dust emission regions in the galaxy, while still
allowing us to resolve variations in the nature of the FIR emission by
using a large number of apertures.  Table~\ref{t2} shows the flux
densities derived in each aperture from \halpha\ to the radio
regime. Note the footnotes to that table, which quantify the effects
of foreground stellar contamination in the short-wavelength data.  We
also investigate the effects of expanding the apertures by a factor of
4 in area for the five brightest dust emission complexes (see further
discussion in \S~\ref{S3.1}).  Table~\ref{t3} presents various
quantities derived from the measurements.  We derive an estimate of the 
3-1000 \mm\ total infrared flux (TIR) using equation 4 of \citet{dale02},
evaluated at the effective MIPS wavelengths:
\begin{equation}     
TIR = (10^{-14})(19.74 \cdot S_{24} + 3.23 \cdot S_{70} + 2.59 \cdot S_{160})
\end{equation}
\noindent where S$_{24}$, S$_{70}$ and S$_{160}$ are the measured flux
densities in units of Jy.  Table~\ref{t3} also presents the
\halpha/TIR ratio, the dust-to-gas ratio (see more detailed discussion
below, and also Tables~\ref{t3} and \ref{t4}), and the ratio of
TIR/radio luminosity as represented by the radio-FIR correlation
coefficient ``q''; each of these quantities will be discussed in more
detail below.

\begin{figure*}[!ht]
\plotone{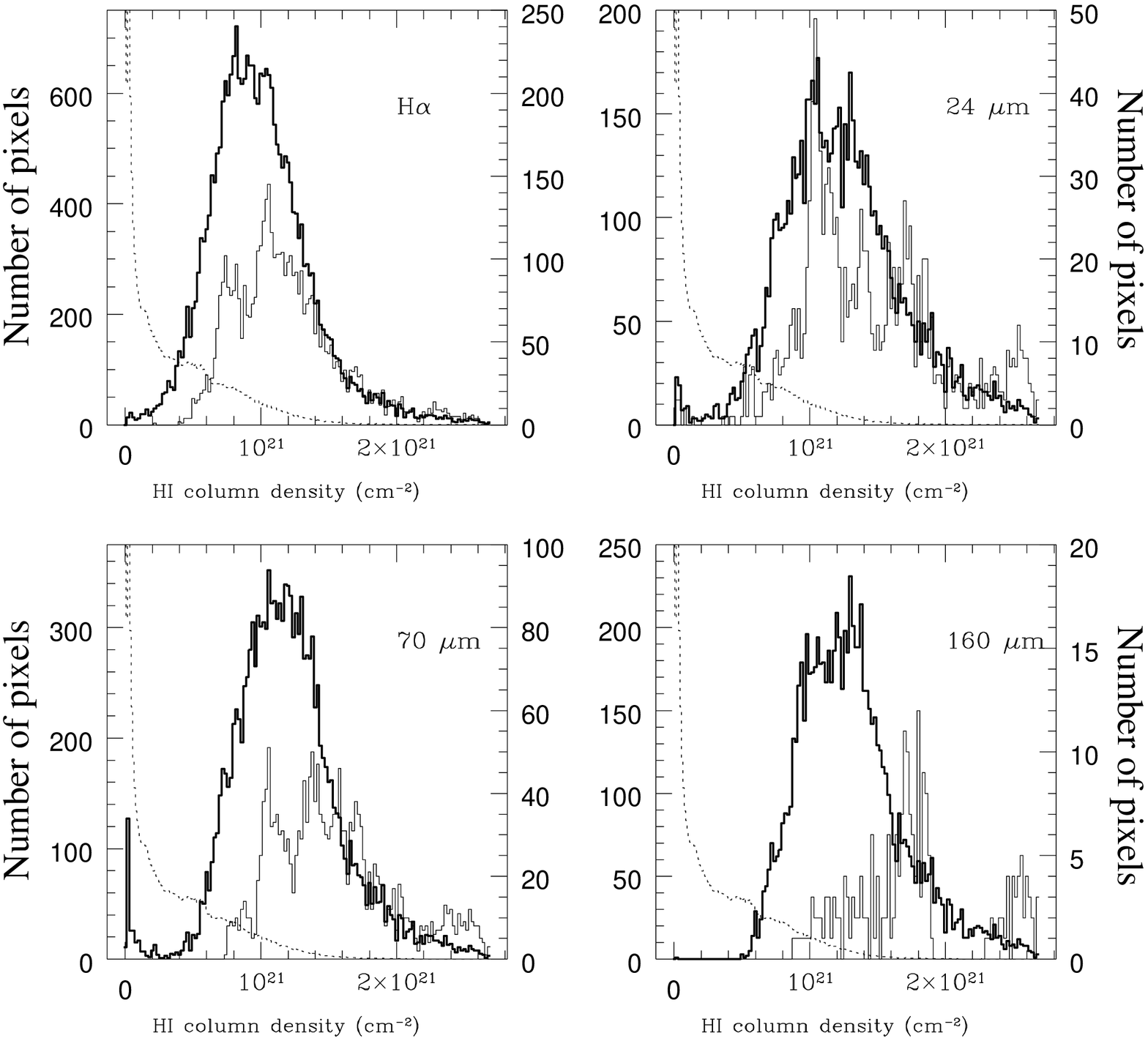}
\caption{Histograms of H~{\sc i} column densities where high-surface
    brightness \halpha\ and FIR emission are detected. The thick black
    lines correspond to flux levels of $\sim$
    3.7\,$\times$\,10$^{-18}$ erg\,s$^{-1}$\,cm$^{-2}$ (\halpha) or to
    surface brightness levels of 4 MJy\,sr$^{-1}$ (IR), and are
    enumerated by the left-hand y-axis of each plot; the thin gray
    lines correspond to flux levels of $\sim$
    1.4\,$\times$\,10$^{-17}$ erg\,s$^{-1}$\,cm$^{-2}$ (\halpha) or to
    surface brightness levels of 16 MJy\,sr$^{-1}$ (IR), and are
    enumerated by the right-hand y-axis of each plot. The dotted line
    shows the histogram of \HI\ column densities throughout the entire
    galaxy, normalized arbitrarily.  These plots demonstrate that 
    nebular and dust emission are strongly correlated with regions that 
    are rich in \HI\ gas.}
\label{figcap7}
\end{figure*}

\begin{figure*}[!ht]
\plotone{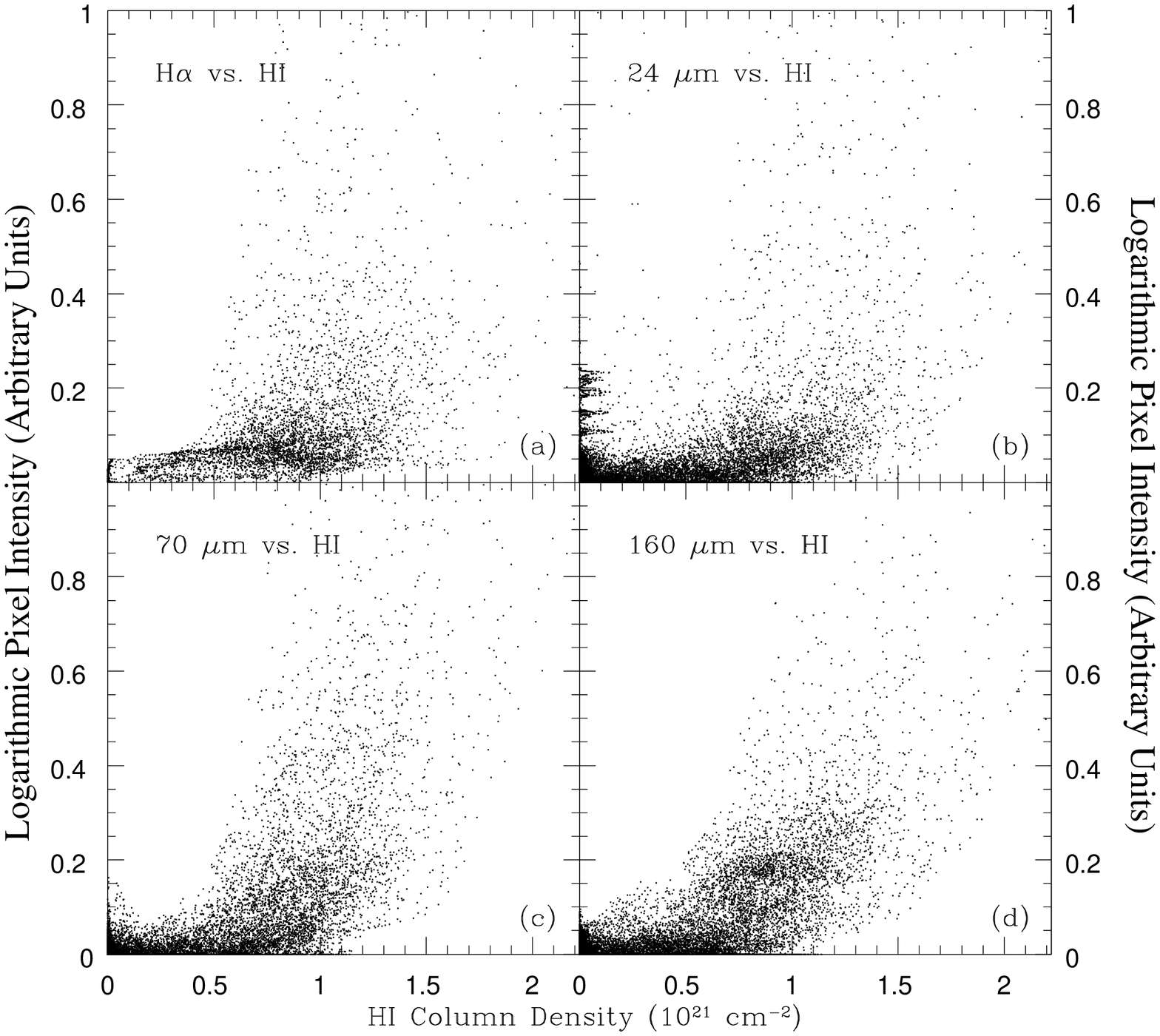}
\caption{Pixel-by-pixel comparison of the strengths of \HI\ and
        \halpha\ (a), 24 \mm\ (b), 70 \mm\ (c), and 160 \mm\ (d)
        emission.  From these plots it is clear that both nebular and
        dust emission is found preferentially in regions of high \HI\
        column density.  The 24 and 70 \mm\ plots (b, c) are
        contaminated by foreground stars and background galaxies at
        low column densities; a first-order estimate of the severity
        of the contamination is available by comparing these
        panels.}
\label{figcap8}
\end{figure*}

\subsection{Linking Heating Sources with the Dust Continuum} 
\label{S3.1}
\subsubsection{\halpha\ vs. IR Emission}
\label{S3.1.1}

The major \HII\ regions Hubble\,I/III, V, and X (regions 3, 8, and 14
in Figure~\ref{figcap3}, respectively) are the strongest \halpha\
sources in the galaxy.  These regions are also luminous at 24, 70 and
160 \mm.  However, the relative strengths of \halpha\ and TIR emission
vary considerably from one source to the next; the regions of
strongest \halpha\ and TIR emission are not always co-spatial. There
are 5 sources in the galaxy with TIR fluxes $\ge$ 1.5 Jy (regions 3,
7, 8, 11, and 14; see Table~\ref{t3}); amongst these regions, the
\halpha/TIR ratio varies by a factor of ten, with region 11 having
strong TIR emission but weak \halpha, and with region 3 having strong
\halpha\ compared to TIR.  Expanding these apertures by a factor of 2
in radius shows the same trends, but with reduced extrema (a factor of
5 variation in the \halpha/TIR ratio is found between regions 11 and
3).  Similar scatter is seen across the entire sample of 16 regions
(see Table~\ref{t2}).

\begin{figure*}[!ht]
\plotone{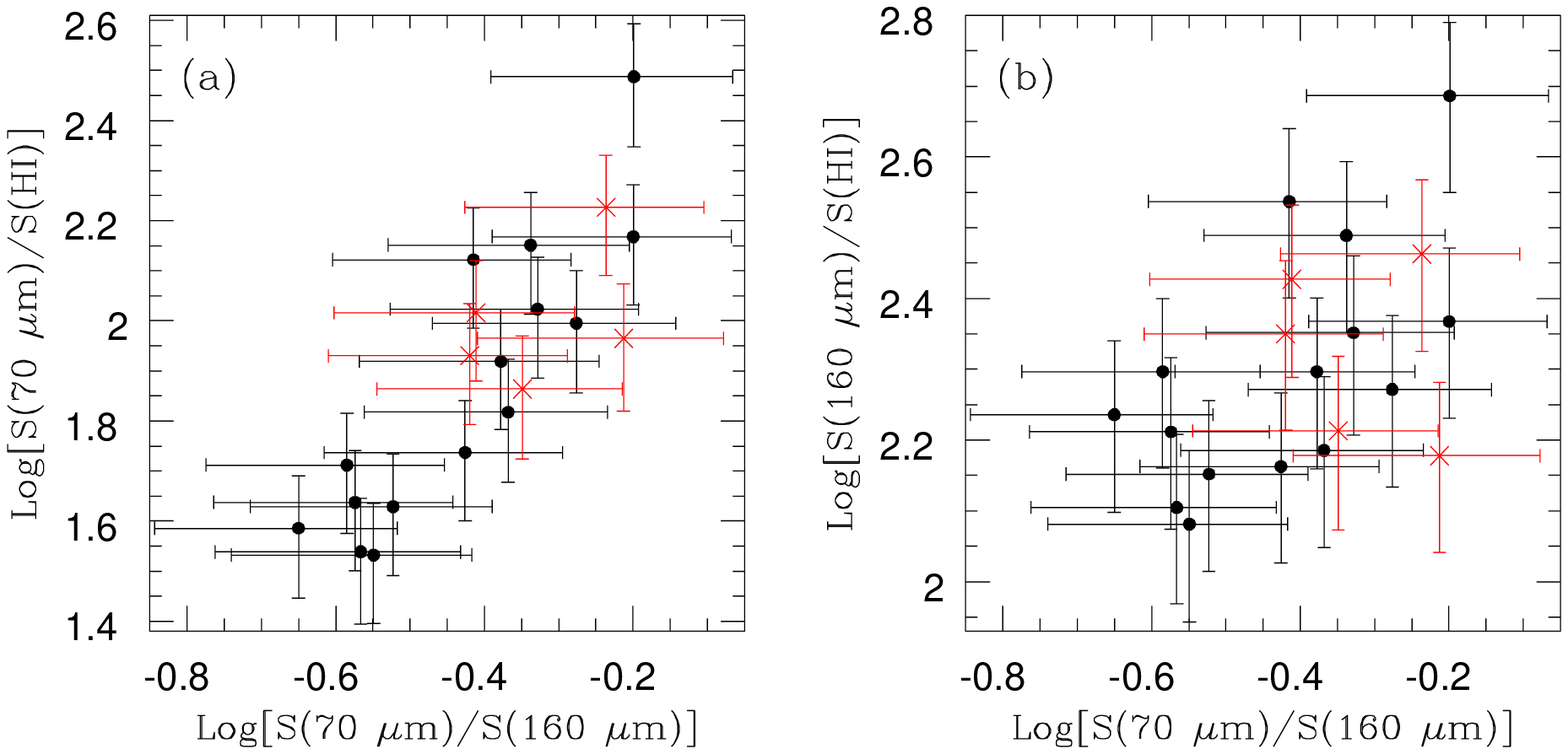}
\caption{Ratios of IR to \HI\ emission strength as a function of dust
        temperature: (a) shows that as the dust temperature increases
        (i.e., higher radiation field strengths), there is more
        70 \mm\ emission per unit \HI\ mass; this suggests that
        local star formation is an important parameter in determining
        the characteristics of IR emission.  (b) shows that cooler
        dust is less sensitive to heating; while a marginal trend
        exists for higher 160 \mm\ fluxes per unit \HI\ mass in
        higher-temperature environments, it is less evident than at 70
        \mm\ (as expected).  Small-radius (57\arcsec) apertures are
        shown as black points, while larger-radius apertures
        (114\arcsec) are shown in red. Note that region 16 is not
        plotted, since the local background is not well-defined.}
\label{figcap9}
\end{figure*}

We now examine what fraction of the observed variations can be
attributed to extinction effects.  Recent work by \citet{wyder03} and
\citet{lee06a} shows that there is variable extinction throughout the
disk, with some regions showing no optical extinction and others
showing A$_{\rm V}$ values up to $\sim$2 magnitudes.  For
strongly star-forming regions with low \halpha/TIR ratios (e.g.,
regions 7 and 11), the \halpha\ flux (and, correspondingly, the
\halpha/TIR ratio) may be up to 6 times larger than observed after
correction for potentially strong internal extinction \citep{lee06a}.
It should be noted, however, that these values were extracted using
1.5\arcsec-wide spectroscopic slits, and thus probe extinction on a
much smaller spatial scale than we are sensitive to with our {\it
Spitzer} data.  Thus, while extinction can play a localized role in
some regions, it cannot account entirely for the observed variation in
the \halpha/TIR ratio.

A second measure of the extinction is available by comparing \halpha\
and radio continuum emission strengths. For this comparison, we assume
that the 20\,cm flux densities shown in Table~\ref{t2} arise from
thermal emission processes.  This assumption is justified for most of
the major \HII\ regions in the galaxy, based on two lines of evidence.
First, multi-frequency radio continuum data using both interferometric
\citep{condon87} and single-dish \citep{klein83,klein86} observations
as compiled in \citet{israel96} show that NGC\,6822 is extraordinarily
weak in the nonthermal radio continuum -- nearly all of the emission
throughout the galaxy appears to be of a thermal origin.  Second,
applying the relations in \citet{caplan86}, and assuming the T$_{\rm
e}$ values calculated in \citet{lee06a}, we can use our observed
(i.e., not extinction-corrected) \halpha\ fluxes to predict lower
limits for the strengths of thermal radio continuum emission.  This
exercise shows that nearly all regions have a small or (in most cases)
negligible nonthermal component: only regions 7 (Hubble IV) and 8
(Hubble V) show thermal fractions less than 90\% ($\sim$50\% and 70\%
in regions 7 and 8, respectively).  This suggests that these two
regions, both active in current star formation (as evidenced by their
strong \halpha\ and IR emission), have produced sufficient numbers of
SNe to accelerate the relativistic electrons that give rise to the
apparent nonthermal components.

We now calculate inferred reddening values, proceeding with the
assumption that thermal emission dominates throughout the galaxy. Of
the six regions detected in the radio continuum, only regions 7 and 8
show potentially strong inferred reddening values (A$_{\rm H\alpha}$
\gsim\ 0.25 magnitudes). Regions 2, 3, 4, and 14 are consistent with
values of A$_{\rm H\alpha} <$ 0.1 magnitude.  Note that if regions 7
and 8 do in fact contain a (relatively) strong synchrotron component,
then the (smaller) thermal fraction of the detected flux densities
puts a yet stronger constraint on the low inferred values of the
reddening.  The smaller inferred extinctions using this approach
compared to the values derived from long-slit spectroscopy
\citep{lee06a} are expected, since the latter data probe smaller-scale
dust concentrations.

To investigate the localized effects of photon absorption and
re-radiation in the IR, the central column of Figure~\ref{figcap4}
presents detailed comparisons of the \halpha\ and IR morphologies.
Here, a low-level \halpha\ flux contour (3.7\,$\times$\,10$^{-18}$
erg\,s$^{-1}$\,cm$^{-2}$) is overlaid on each of the 24, 70 and 160
\mm\ images.  It is immediately clear that the regions of high
\halpha\ and IR surface brightnesses are directly linked, since the
\halpha\ contours surround the IR emission peaks. We are thus
observing localized dust heating in the ISM (see also further
discussion below).  However, it is also interesting to note that a
significant fraction of the integrated IR flux density of the galaxy
is unassociated with regions that are luminous at \halpha.  Most
\halpha-bright regions are also luminous in the IR; one
notable exception is region 16 (see Figure~\ref{figcap3}), which shows
(low-level) \halpha\ emission in a clear shell morphology but which
lacks strong associated dust emission.   However, we
treat region 16 with caution, since the local background is not
well-defined (see Table~\ref{t2}) and may be over-subtracted in the 
IR images.

To discern the origin of the photons that give rise to IR emission
throughout the galaxy, it is useful to quantify what fraction of the
total IR luminosity of NGC\,6822 arises from regions that are not
luminous at \halpha.  The global flux densities in the long-wavelength
bands, measured over the entire central disk region (S$_{\rm 24} =$
2.51\,$\pm$\,0.50 Jy; S$_{\rm 70} =$ 53.2\,$\pm$\,15 Jy; S$_{\rm 160}
=$ 136.2\,$\pm$\,40 Jy; see Figure~\ref{figcap3} and Table~\ref{t2}),
appear to consist of discrete FIR peaks and a more extended
``diffuse'' component. Here we classify emission as diffuse if it is
located outside the regions detected at \halpha\ (shown as the
3.7\,$\times$\,10$^{-18}$ erg\,s$^{-1}$\,cm$^{-2}$ contour in
Figure~\ref{figcap4}).  The quantification of the fraction of the
total FIR that arises in a diffuse component is dependent on the
sensitivity of the \halpha\ data to which we compare the IR images
(or, equivalently, on the surface brightness below which we classify
IR emission as ``diffuse'').  The spatial scales over which the
diffuse emission are found are much larger than the PSFs at 160 and
(especially) 70 \mm\ (precluding PSF smearing as the only origin of
the diffuse component).  We find that $>$ 60\% of the integrated flux
densities at 70 and 160 \mm\ arise from this diffuse component.  While
the 24 \mm\ emission is more strongly peaked in regions of active star
formation \citep[e.g.,][]{helou04,calzetti05}, the diffuse component
still accounts for $\sim$ 50\% of the total 24 \mm\ flux
density. These results suggest that a substantial amount of the IR
radiation may be powered by non-ionizing sources or may be the result
of UV photons escaping directly from star formation regions; this is
discussed further in \S~\ref{S4}.  Similar results were obtained for
NGC\,6822 by \citet{gallagher91} and \citet{israel96}; diffuse far-IR
emission is also seen in other nearby dwarf galaxies \citep[e.g.,
NGC\,55;][]{engelbracht04}.

In Figure~\ref{figcap5} we present elliptical surface brightness
profiles of \HI, \halpha\ and IR emission. These profiles were derived
using parameters from the \HI\ rotation curve analysis of
\citet{deblok00} and \citet{weldrake03}.  Radially-averaged elliptical
annuli with 12\arcsec\ thickness are integrated along the position
angle of the galaxy's major axis [central position: $\alpha =$
19:44:58.04, $\delta = -$14:49:18.9 (J2000; this location is between
apertures 9 and 11 in Figure~\ref{figcap3}); position angle $\simeq$
120\degree; see {Weldrake \etal\ 2003}\nocite{weldrake03} for
details].  The resulting surface brightness profiles show that
\halpha, IR and \HI\ peaks correlate remarkably well.  In particular,
the \halpha\ and 24 \mm\ profiles are very similar, tracing the local
star formation rate ({Helou \etal\ 2004}\nocite{helou04}; {Calzetti
\etal\ 2005}\nocite{calzetti05}).  We discuss the surface brightness
profiles in more detail in \S~\ref{S3.2}.

\subsubsection{Infrared vs. Optical Luminosities and Star Formation Rates}
\label{S3.1.2}

The metal-poor ISM of NGC\,6822 is an interesting environment in which
to compare the luminosities and implied SFRs from various methods.
Using the global flux densities in the long-wavelength bands (see
above) and the relations presented in \citet{dale02}, the total IR
flux of $\sim$ 5.7\,$\times$\,10$^{-12}$ W\,m$^{-2}$ corresponds to
L$_{\rm TIR} \simeq$ 1.6\,$\times$\,10$^{41}$ erg\,s$^{-1}$ at 490
kpc.  The \halpha-based SFR (sensitive to recent star formation over
the last $\sim$ 10 Myr) is $\sim$ 0.016 \msun\,yr$^{-1}$, and the
total recent SFR (over the last $\sim$ 100-200 Myr) derived from the
stellar population study of \citet{gallart96b} is $\sim$0.04
\msun\,yr$^{-1}$.

The 24 \mm\ luminosity (L$_{\rm 24}$), which has been shown to
correlate well with regions of active star formation in a range of
environments \citep[e.g., ][]{helou04,calzetti05}, is an appropriate
IR-based SFR indicator for dust-rich regions.  We compare the L$_{\rm
24}$/\halpha\ ratio for NGC\,6822 with what was observed by
\citet{calzetti05} in the more metal-rich spiral galaxy M\,51
\citep{bresolin04}, as follows:
\begin{equation}     
Log(L_{24}) = 1.03Log({L_{H\alpha}}) - 0.06945
\end{equation}
where L$_{\rm 24}$ and L$_{\rm H\alpha}$ are the observed luminosities
at 24 \mm\ and \halpha, respectively; the intrinsic \halpha/P$\alpha$
ratio is assumed to be 8.734.  The central wavelength of 24.0 \mm\ is
used to convert the 24 \mm\ flux density to a luminosity measurement
[thus, L$_{\rm 24}$ as defined above is not strictly a luminosity, but
is rather equivalent to $\nu$\,$\cdot$\,F$_{\nu}$; we keep the above
notation to allow direct comparison to the work of {Calzetti \etal\
(2005)}\nocite{calzetti05}].  Using the total observed \halpha\
luminosity of the system (see Table~\ref{t2}), this relation predicts
a total 24 \mm\ luminosity $\sim$3 times higher than the observed
value.  This confirms that metal-poor galaxies do not have enough
metals to form dust in the same fraction as more metal-rich
environments.

\begin{deluxetable*}{lccccccc}[ht]
\tabletypesize{\scriptsize}
\tablecaption{Derived Properties of NGC\,6822} 
\tablewidth{0pt} 
\tablehead{
\colhead{Region\tablenotemark{a}} 
&\colhead{S(TIR)\tablenotemark{b}}
&\colhead{F(\halpha)/F(TIR)\tablenotemark{c}} 
&\colhead{\HI\ Mass}
&\colhead{M$_{\rm Dust}$/M$_{\rm HI}$\tablenotemark{d}} 
&\colhead{q$_{\rm TIR}$\tablenotemark{e}}
&\colhead{q$_{70}$\tablenotemark{e}}
&\colhead{q$_{24}$\tablenotemark{e}}\\ 
&(10$^{-13}$ W\,m$^2$) & &(10$^5$ \msun) &(\dgr) & & &} 
\startdata
\multicolumn{8}{c}{Aperture radii $=$ 57\arcsec\ $\simeq$ 3\,$\times$\,(160 \mm\ PSF FWHM $\simeq$ 130 pc)}\\
\hline 
1 &1.03\pom0.41  &9.6\pom4.3 &7.3\pom0.8  &0.0038\pom0.0021    &\nodata &\nodata &\nodata\\ 
2 &0.73\pom0.29  &36\pom16   &6.9\pom0.7  &0.0045\pom0.0024    &2.76    &2.15    &0.78\\ 
3 &1.98\pom0.79  &55\pom24   &6.8\pom0.7  &0.0048\pom0.0026    &2.66    &2.15    &0.96\\ 
4 &0.85\pom0.34  &28\pom13   &8.4\pom0.9  &0.0039\pom0.0021    &2.87    &2.31    &1.01\\ 
5 &0.80\pom0.32  &6.3\pom2.8 &6.2\pom0.7  &0.0034\pom0.0018    &\nodata &\nodata &\nodata\\ 
6 &0.76\pom0.30  &18\pom8    &5.7\pom0.7  &0.0046\pom0.0025    &\nodata &\nodata &\nodata\\ 
7 &2.88\pom1.15  &8.3\pom3.8 &12.4\pom1.2 &0.0037\pom0.0020    &3.08    &2.56    &1.61\\ 
8 &5.39\pom2.16  &23\pom10   &7.9\pom0.8  &0.0073\pom0.0039    &2.83    &2.31    &1.57\\ 
9 &0.91\pom0.36  &5.1\pom2.2 &7.8\pom0.8  &0.0026\pom0.0014    &\nodata &\nodata &\nodata\\ 
10 &0.69\pom0.28 &7.0\pom3.2 &5.2\pom0.6  &0.0020\pom0.0010    &\nodata &\nodata &\nodata\\ 
11 &2.40\pom0.96 &5.4\pom2.5 &8.6\pom0.9  &0.0066\pom0.0035    &\nodata &\nodata &\nodata\\ 
12 &0.52\pom0.21 &11\pom5    &6.2\pom0.7  &0.0022\pom0.0012    &\nodata &\nodata &\nodata\\ 
13 &0.70\pom0.28 &9.7\pom4.4 &8.0\pom0.8  &0.0030\pom0.0016    &\nodata &\nodata &\nodata\\ 
14 &2.42\pom0.97 &46\pom20   &9.2\pom1.0  &0.0025\pom0.0013    &2.75    &2.34    &1.24\\ 
15 &1.07\pom0.43 &15\pom7    &6.0\pom0.6  &0.0035\pom0.0019    &\nodata &\nodata &\nodata\\ 
16\tablenotemark{f} &0.29\pom0.12 &59\pom28   &6.6\pom0.7  &0.00055\pom0.00029  &\nodata &\nodata &\nodata\\ 
\hline
\multicolumn{8}{c}{Aperture radii $=$ 114\arcsec\ $\simeq$ 6\,$\times$\,(160 \mm\ PSF FWHM) $\simeq$ 260 pc }\\ 
\hline 
3  &4.0\pom1.6 &33\pom14   &23.3\pom2.4  &0.0050\pom0.0027  &2.97        &2.50    &1.19\\ 
7  &5.0\pom2.0 &8.4\pom3.7 &36.4\pom3.7  &0.0030\pom0.0016  &3.01        &2.56    &1.40\\ 
8  &7.8\pom3.1 &19\pom8.6  &25.6\pom2.6  &0.0040\pom0.0021  &2.99        &2.54    &1.62\\ 
11 &5.7\pom2.3 &6.1\pom2.8 &28.4\pom2.9  &0.0048\pom0.0026  &\nodata &\nodata &\nodata\\
14 &4.1\pom1.6 &32\pom15   &30.1\pom3.0  &0.0016\pom0.00087 &2.80        &2.46    &1.18\\ 
\hline
\multicolumn{8}{c}{Total Galaxy}\\ 
\hline 
NGC\,6822 &57\pom23 &12.3\pom5.5 &1340\pom140 &0.00077\pom0.00043\tablenotemark{g} &3.34 &2.88 &1.55\\ 
\enddata
\label{t3}
\tablenotetext{a}{See Table 1 for source coordinates and cross-identifications.}
\tablenotetext{b}{Calculated using the relations presented in \citet{dale02}; uncertainties are estimated at the $\sim$ 40\% level.}\\
\tablenotetext{c}{Calculated as the unitless ratio of \halpha\ to TIR flux; ratios represent 1000\,$\times$ this fraction.}
\tablenotetext{d}{See Table~\ref{t4} for dust mass derivations; the values shown here are calculated using dust masses derived from the models of \citet{draine01}, \citet{li01,li02}, and \citet{draine06}.}
\tablenotetext{e}{q$_{\rm TIR}$ is the ``q'' parameter evaluated using an estimate of the total IR luminosity (log(S(TIR/Radio))); q$_{24}$ and q$_{70}$ are the monochromatic ``q'' parameters, as studied in {Murphy \etal\ (2006)}.}
\tablenotetext{f}{Note that the local background surrounding region \#16 in the MIPS images is very uncertain; see \S~\ref{S3.1.1}.}
\tablenotetext{g}{This value uses the total \HI\ mass from the entire system; using the \HI\ mass contained in the matching aperture used for the global IR flux density extractions reduces the enclosed \HI\ mass by $\sim$ 10\%.}
\end{deluxetable*}

\begin{deluxetable*}{lccccc}
\tabletypesize{\scriptsize}
\tablecaption{Dust Mass Derivations in NGC\,6822}
\tablewidth{0pt}
\tablehead{
\colhead{Region\tablenotemark{a}} &\colhead{T$_{\rm DUST}$\tablenotemark{b}} &\colhead{Blackbody} &\colhead{$\alpha$\tablenotemark{c}} &\colhead{Dale \etal} &\colhead{Li \& Draine}\\
&\colhead{(K)} &\colhead{Dust Mass\tablenotemark{b}} &\colhead{} &\colhead{Dust Mass\tablenotemark{c}}  &\colhead{Dust Mass\tablenotemark{d}}\\
&\colhead{} &\colhead{(10$^3$ \msun)} &\colhead{} &\colhead{(10$^3$ \msun)}  &\colhead{(10$^3$ \msun)}}
\startdata
\multicolumn{6}{c}{Aperture radii $=$ 57\arcsec\ $=$ 3\,$\times$\,(160 \mm\ PSF FWHM \gsim\ 130 pc)}\\\hline
1  &21\pom2 &1.4\pom0.7     &2.03 &12.6\pom6.3   &2.81\pom1.40\\   
2  &21\pom2 &1.6\pom0.8     &2.88 &14.4\pom7.2   &3.11\pom1.55\\   
3  &24\pom2 &1.6\pom0.8     &2.31 &14.4\pom7.2   &3.27\pom1.63\\   
4  &22\pom2 &1.3\pom0.6     &2.75 &11.7\pom5.9   &3.26\pom1.63\\   
5  &23\pom2 &0.82\pom0.41   &2.00 &7.4\pom3.7    &2.11\pom1.05\\   
6  &21\pom2 &1.4\pom0.7     &2.81 &12.6\pom6.3   &2.63\pom1.31\\   
7  &24\pom2 &2.1\pom1.1     &2.00 &18.9\pom9.5   &4.62\pom2.31\\   
8  &25\pom2 &2.3\pom1.1     &1.72 &20.7\pom10.4  &5.79\pom2.89\\   
9  &23\pom2 &0.97\pom0.49   &2.62 &8.7\pom4.4    &2.04\pom1.02\\   
10 &24\pom2 &0.66\pom0.33   &2.50 &5.9\pom3.0    &1.01\pom0.50\\   
11 &23\pom2 &2.6\pom1.3     &2.56 &23.4\pom11.7  &5.61\pom2.80\\   
12 &21\pom2 &0.93\pom0.47   &2.75 &8.4\pom4.2    &1.40\pom0.70 \\  
13 &21\pom2 &1.1\pom0.6     &2.81 &9.9\pom5.0    &2.43\pom1.21\\   
14 &25\pom2 &1.3\pom0.6     &2.03 &11.7\pom5.9   &2.31\pom1.15\\   
15 &23\pom2 &0.98\pom0.49   &2.19 &8.8\pom4.4    &2.09\pom1.04\\   
16\tablenotemark{e} &24\pom2 &0.23\pom0.12   &2.50 &20.7\pom10.4  &0.36\pom0.18\\ 
\hline
\multicolumn{6}{c}{Aperture radii $=$ 114\arcsec\ $=$ 6\,$\times$\,(160
  \mm\ PSF FWHM) \gsim\ 260 pc }\\
\hline
3  &23\pom2 &4.7\pom2.3  &2.56 &42.3\pom21.2 &11.7\pom5.9\\
7  &24\pom2 &4.7\pom2.3  &2.28 &42.3\pom21.2 &11.1\pom5.5\\
8  &25\pom2 &4.7\pom2.3  &1.91 &42.3\pom21.2 &10.2\pom5.1\\
11 &23\pom2 &6.7\pom3.3  &2.62 &60.3\pom30.2 &13.6\pom6.8\\
14 &25\pom2 &2.8\pom1.4  &2.25 &25.2\pom12.6 &4.91\pom2.4\\
\hline
\multicolumn{6}{c}{Total Galaxy}\\
\hline
NGC\,6822 &23\pom2 &83\,\pom\,42 &2.56 &750\pom380  &103\pom52\\
\enddata
\label{t4}
\tablenotetext{a}{See Table 1 for source coordinates and cross-identifications.}
\tablenotetext{b}{Average dust temperature and implied dust mass over the aperture, derived by blackbody fitting.}
\tablenotetext{c}{$\alpha$ and dust mass values derived using the semiempirical SED models of \citet{dale01} and \citet{dale02}; typical errors on the $\alpha$ index are $\sim$ 0.15 (see discussion in \S~\ref{S3.2.2}).}
\tablenotetext{d}{Dust mass derived using the SED models of \citet{draine01}, \citet{li01,li02}, and \citet{draine06}.}
\tablenotetext{e}{Note that the local background surrounding region \#16 in the IR images is very uncertain; see \S~\ref{S3.1.1}.}
\end{deluxetable*}

To explore the relation between \halpha\ and IR emission on a
spatially resolved basis, we plot in Figure~\ref{figcap6} the \halpha\
vs. 24 \mm\ fluxes, and compare to the relation derived for M\,51.
While there is an appreciable scatter, it is clear that regions within
NGC\,6822 follow a different scaling between \halpha\ and 24 \mm\
luminosity than the regions in M\,51: the \HII\ regions in NGC\,6822
have much lower extinction than those in M\,51, as expected on the
basis of the widely different metallicities (factor $\sim$10-15; see
{Bresolin \etal\ 2004}\nocite{bresolin04}).  Note that extinction
corrections at \halpha\ will move points to the right in this plot,
and cannot directly account for all of the observed dispersion (see
further discussion in \S~\ref{S3.1.1}); we plot the largest detected
reddening vector (0.75 mag, for aperture 7 or Hubble\,IV) in
Figure~\ref{figcap6} to demonstrate the severity of (spatially
variable) extinction at \halpha.  The scatter in this plot can then be
interpreted as the effect of star formation on the local ISM. This
expands the findings of \citet{cannon05,cannon06}, where the
vigorously star-forming, metal-poor galaxies IC\,2574 and NGC\,1705
were also found to have different \halpha/24 \mm\ ratios than those
found in more metal-rich galaxies.  Further, it reiterates the conclusions
of \citet{kennicutt98}: the infrared luminosity is a reliable SFR
indicator only in metal-rich (i.e., dusty) environments, where a large
fraction of the light from massive stars is reprocessed by dust to the
infrared. We discuss these results further in \S~\ref{S4}.

\subsection{Neutral Gas vs. Dust Emission}
\label{S3.2}
\subsubsection{Correlations Between \HI\ and Dust}
\label{S3.2.1}

The right column of Figure~\ref{figcap4} shows the IR images overlaid
with contours of \HI\ column density at the 10$^{21}$ cm$^{-2}$
level. It is clear from this image and Table~\ref{t2} that the central
disk of NGC\,6822 is rich in neutral gas and has relatively small
variations in \HI\ column density over the apertures in this study
(factor of $\sim$ 2 variations in flux).  Comparing the locations of
\halpha, \HI\ and dust emission in Figure~\ref{figcap4}, it is also
clear that nearly all locations of active star formation (as evidenced
by high-surface brightness \halpha\ emission) are associated with \HI\
columns in excess of $\sim$10$^{21}$ cm$^{-2}$.  Note by comparison
with Figures~\ref{figcap1} and \ref{figcap2}, however, that the \HI\
distribution is much more extended than the high-surface brightness
optical body and the areas with strong FIR emission.

The surface brightness profiles of the FIR emission (see
Figure~\ref{figcap5}) show that the 24 \mm\ emission is, in general,
associated with local \HI\ column density maxima.  Moving toward
cooler dust emission, the 70 and 160 \mm\ profiles show a much lower
dynamic range than the 24 \mm\ profile; however, the peaks in the FIR
nicely correlate with the locations of \HI\ surface density maxima.
For example, at radii of $\sim$0.2, 1.0, and 1.6 kpc, the correlations
between \HI, \halpha\ and IR emission are especially strong (though
relative variations are evident).

\begin{figure*}[!ht]
\plotone{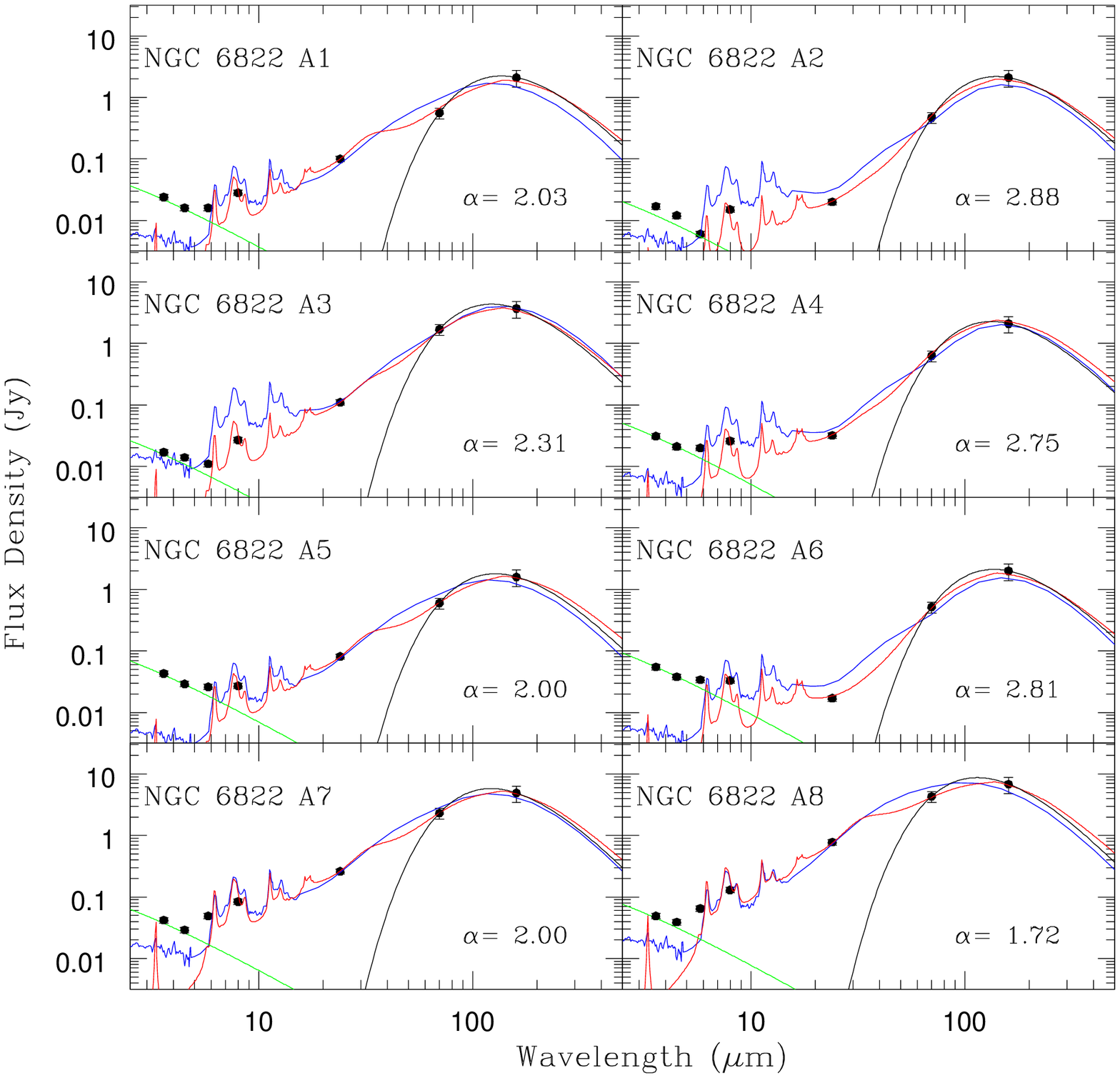}
\caption{Modified blackbody and SED model fits to the observed data
  for apertures A1--A8 (see Table~\ref{t1} and Figure~\ref{figcap3}).
  Each panel shows the observed infrared flux densities, overlaid
  with: a single-temperature modified blackbody fit to the observed 70
  and 160 \mm\ flux densities (fits shown in black); the SED models of
  {Dale \& Helou (2002)}, fitting to the observed 24, 70 and 160 \mm\
  flux densities (fits shown in blue); the models of {Li \& Draine
  (2001, 2002)} and {Draine \& Li (2006)}, fitting to
  all observed flux densities longward of 5 \mm\ (fits shown in red);
  and a stellar component based on population synthesis models, fit to
  the 3.6 and 4.5 \mm\ bands (green line).  Note that
  single-temperature modified blackbody fits strongly underestimate
  the observed 24 \mm\ flux densities in all cases.  The value of
  $\alpha$ from the {Dale \& Helou (2002)} fits is shown at the bottom
  right of each panel.}
\label{figcap10}
\end{figure*}

\begin{figure*}[!ht]
\plotone{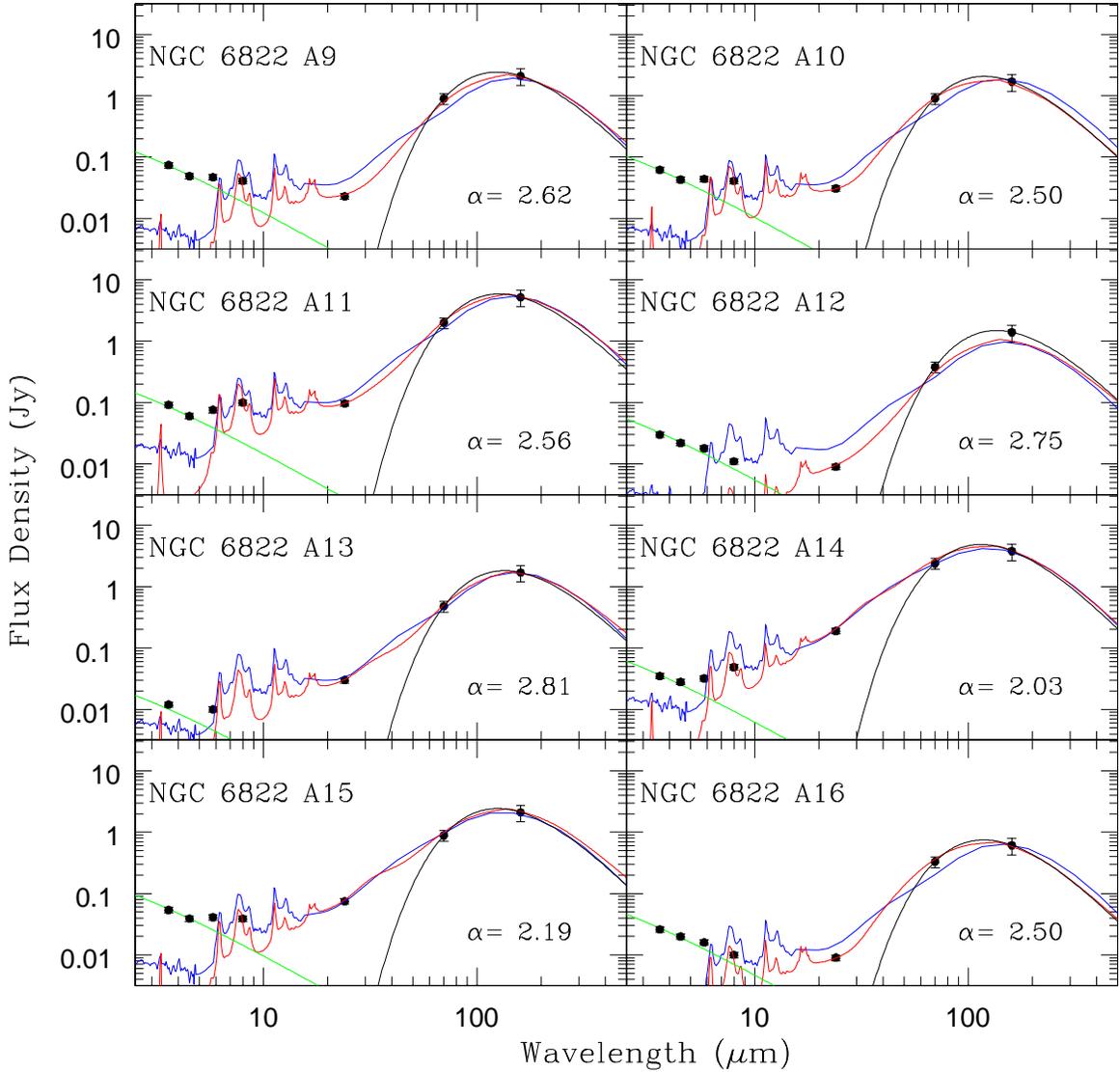}
\caption{Same as Figure~\ref{figcap10}, but for apertures A9--A16.}
\label{figcap11}
\end{figure*}

\begin{figure*}[!ht]
\plotone{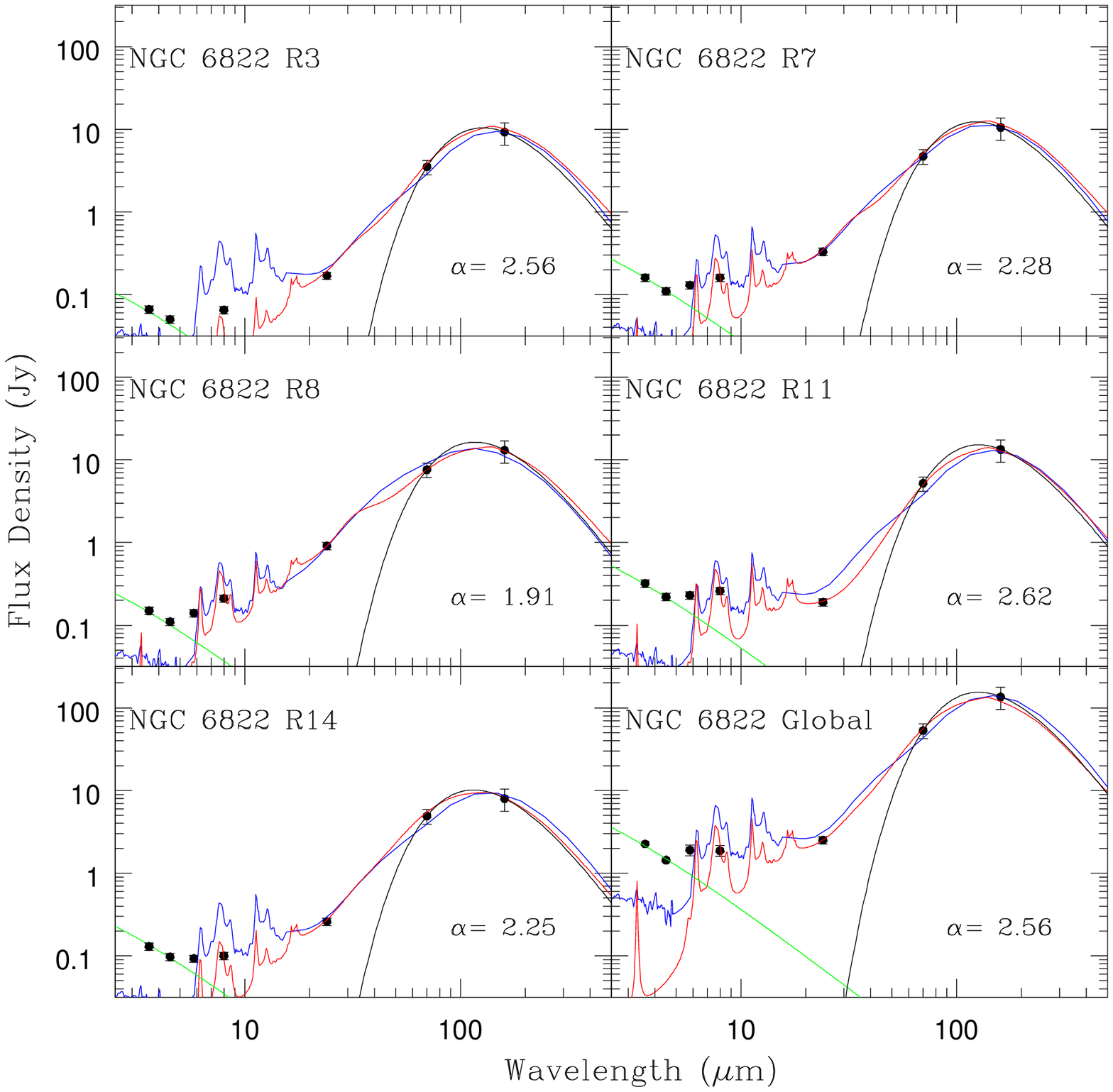}
\caption{Same as Figure~\ref{figcap10}, but for the apertures with
     expanded radii (see Table~\ref{t2}) and for the global SED of
     NGC\,6822.  Note that the y-axis has been increased compared to
     Figures~\ref{figcap10} and \ref{figcap11}.}
\label{figcap12}
\end{figure*}

To further quantify the correlation between \HI\ and FIR emission,
Figures~\ref{figcap7} and \ref{figcap8} show pixel-by-pixel
histograms and scatter plot comparisons of \HI\ column density
vs. surface brightness measurements in the \halpha\ and IR bands.
These plots were created by comparing \HI, \halpha, and IR surface
brightnesses on an individual pixel basis. They clearly demonstrate
that FIR and \halpha\ emission are strongly peaked in regions with
\HI\ column densities $\gsim$ 10$^{21}$ cm$^{-2}$.  The 70 and 160
\mm\ histograms show that the regions of strongest dust emission (as
traced by very high-surface brightness 70 and 160 \mm\ emission; thin
gray lines in Figure~\ref{figcap7}) are associated with the highest
\HI\ column densities: the histogram at 16 MJy\,sr$^{-1}$ shows a
median value $\sim$ 50\% higher than the median at 4
MJy\,sr$^{-1}$. There is little difference in the pixel-by-pixel
distribution of surface brightnesses at \halpha\ or in the IR bands;
above \HI\ columns of $\sim$5\,$\times$\,10$^{20}$, the nebular and
dust emission both increase in strength as the \HI\ column increases,
with the highest-surface brightness emission strongly peaked in regions 
with \HI\ column densities $\gsim$10$^{21}$ cm$^{-2}$.

We now consider the origin of the evident correlation between nebular
and IR surface brightness as a function of \HI\ column density.  If
the strength of the dust emission is correlated primarily with dust
heating, then the far-IR colors should become warmer as the ratio of
dust to \HI\ emission increases.  However, if the strength of the dust
emission is related to an increase in the density of dust, then the
ratio of infrared to \HI\ emission will not be correlated with the
far-infrared colors.  To investigate this scenario, we plot in
Figure~\ref{figcap9} the ratios of S$_{\rm 70}$/S$_{\rm HI}$ and
S$_{\rm 160}$/S$_{\rm HI}$ for the regions listed in Table~\ref{t1},
each as a function of IR color [here, Log(S$_{\rm 70}$/S$_{\rm
160}$)].  These plots allow us to gauge how much IR emission is
arising per unit \HI\ mass at a given dust temperature (or, similarly,
at a given radiation field strength level).  It is clear
from Figure~\ref{figcap9}(a) that as the IR color increases (radiation
field strength rising), the amount of 70 \mm\ emission per unit \HI\
mass also increases.  At 160 \mm, this correlation is less evident,
suggesting that 70 \mm\ emission is more sensitive to local heating,
as one would expect for thermal emission in the temperature range of
interstellar dust \citep[see also][]{dale05}.  This arises in part due to the 70
\mm\ band falling near the transition point between the hot dust
emission that dominates shorter wavelengths and the $\sim$20--30 K
dust emission that radiates at longer wavelengths.  While the
errorbars in both plots are large, we can infer that localized heating
of dust, in addition to increased \HI\ column density, plays an
important role in controlling the nature of IR emission.  The different
dynamic ranges of nebular, dust and \HI\ emission may also suggest 
increasing \HI\ columns due to photodissociation of H$_2$ in regions of 
active star formation.

The simplest physical scenario that explains these observational
trends is one where the \HI\ and dust are mixed in the ISM, and that
as the local radiation field strength increases, so does the total IR
emission per unit dust mass \citep[e.g.,][]{dale99}.  If the increase in the dust emission
surface brightness was due solely to an increase in the local \HI\
column density (i.e., a relatively constant-strength interstellar
radiation field producing higher surface brightnesses in regions of
larger \HI\ columns), then the dynamic range of the dust emission
would be expected to be similar to that in \HI.  However,
Figures~\ref{figcap7} and \ref{figcap8} clearly show a large scatter
of IR flux densities at a given \HI\ column or a given radiation field
strength.  We caution that with the present observations, it is not
possible to assure that the \HI, \halpha\ and IR observations are all
sampling interstellar media that are exactly co-spatial; however, the
clear correlations between these various phases (see
Figures~\ref{figcap4}, \ref{figcap7} and \ref{figcap8}) suggest that
dust absorption and re-radiation are closely tied to local star
formation intensity.  Note that no regions with high IR surface
brightness and high \HI\ column density are \halpha-deficient (i.e.,
non-detections); this again supports the theory that local star
formation is an important mechanism in the production of the observed
IR SED.

\subsubsection{Derivation of Dust Masses}
\label{S3.2.2}

In the following, we calculate the dust mass of NGC\,6822 (both
globally and within the apertures shown in Figure~\ref{figcap3}) using
three different methods of increasing sophistication.  We begin with a
single-temperature fit, and discuss explicitly the shortcomings of
such an approach in accurately representing the observed SEDs within
NGC\,6822.  We then discuss two SED modeling techniques that offer a
more physical representation of the data.

First, we apply a simple single-temperature modified blackbody fit to
the data \citep[see, e.g.,][]{hildebrand83,bianchi99}.  Dust mass
values are derived via the following equation:
\begin{equation}     
M_{dust} = \frac{D^2 f_{\nu}}{\kappa_{\nu} B(T)}
\end{equation}
where $D$ is the distance to the galaxy, $f_{\nu}$ is the flux
density, $\kappa_{\nu}$ represents the absorption opacity of the dust
(given in {Li \& Draine 2001\nocite{li01}), and $B(T)$ is the Planck
function evaluated at temperature T, derived by fitting a blackbody
modified by a $\lambda^{-2}$ emissivity function to the 70 and 160
\mm\ data.  Using this method, we derive a total dust mass of
(8.3\,\pom\,4.2)\,$\times$\,10$^{4}$ \msun; Table~\ref{t4} shows the
inferred dust masses for each region.

This simple prescription assumes that all the IR emission in a galaxy
is at both the same temperature and in thermal equilibrium.  A
combination of the composite nature of IR emission [arising from large
grains, polycyclic aromatic hydrocarbons (PAH), very small grains,
etc.], the stochastic heating processes that are of differing
importance for various types of grains (see {Draine \& Li
2001}\nocite{draine01} for a detailed discussion), and the complex
relations between dust emission, gas density, and the local
interstellar radiation field,, suggests that this single-temperature
assumption is likely incorrect, at least on local scales.  Indeed,
Figures~\ref{figcap10}, \ref{figcap11} and \ref{figcap12} show that the
simple blackbody fits to the 70 and 160 \mm\ data are not intended to
represent the hot dust emission at shorter wavelengths; the fits fall
short of the observed 24 \mm\ data points by orders of magnitude.
Given these failures, we seek a more detailed
treatment of the SEDs that accounts for the potential
contribution of multiple grain populations to the dust mass.

A second and more robust method with which to calculate the dust mass
using IR to submillimeter data is with the SED models presented by
\citet{dale01} and \citet{dale02}.  These models assume that large
grains, very small grains, and PAH molecules contribute to the
integrated IR energy budget in a galaxy. Semiempirical SEDs are
constructed assuming a power-law distribution of dust mass as a
function of the strength of the local interstellar radiation field
(U).  With U normalized to unity for the radiation field near the sun,
masses are computed via the relation:
\begin{equation}
dM_{dust} (U) \propto U^{-\alpha} dU
\end{equation}
where M$_{\rm dust}$ is the dust mass heated by a radiation field of strength
$\leq$ U, and $\alpha$ represents the relative contributions of the different local
SEDs. The fitted values of $\alpha$ correlate with an empirical measure of the
ratio \begin{math}{\rm Log}[f_{\nu}(60 \mu m)/f_{\nu}(100 \mu m)]\end{math},
which was calibrated against \iras\ data \citep[see][]{dale01,dale02}.  This
ratio then corresponds to a scaling factor by which the dust mass estimates
using single-temperature blackbody fits (see above) should be increased to
account for mass components not well-constrained by the blackbody fit (see
application of this method to NGC\,7331 in {Regan \etal\
2004}\nocite{regan04}).  Values of $\alpha$ between $\sim$ 2--4 imply a mass
increase of a factor of 9 (see Figure~6 of {Dale \& Helou
2002}\nocite{dale02}); $\alpha$ values less than this show a smaller increase
in dust mass (e.g., $\alpha =$ 1.75 corresponds to a factor of $\sim$ 8 
increase).

Fitting these models to the flux densities of the regions and the entire
galaxy (see Tables~\ref{t2} and \ref{t4}) yields values of 2.0 $\lsim\
\alpha\ \lsim$2.8 for nearly all regions; Monte Carlo simulations of the
errors on the index $\alpha$ were performed during the fitting of these models
to 68 galaxies in \citet{dale05}, providing typical errors of $\sim$ 0.15.
The observed values of $\alpha$ in NGC\,6822 are consistent with those seen in
the {Dale \etal}\nocite{dale05} sample; the only exception is region 8
(Hubble\,V), the strongest IR, \halpha\ and radio continuum source in the
galaxy.  This active star formation region may have a strong enough local
radiation field that a large component of cool grains (which would lead to a
pronounced mass increase) is absent.  Figures~\ref{figcap10}, \ref{figcap11}
and \ref{figcap12} show these model fits to each region, and Table~\ref{t4}
gives the values of $\alpha$ for each fit and the inferred dust mass using
these SED models.  Given the uncertainties in the dust masses derived from the
blackbody fits (see above and Table~\ref{t4}) and the varying quality of the
least-squares fit for each region (compare, e.g., the fits for regions 6 and
14), all regions are consistent with an increase in mass by a factor of $\sim$
9; the global dust mass increases accordingly to
(7.5\,\pom\,3.8)\,$\times$\,10$^{5}$ \msun.

A third and yet more sophisticated estimate of the dust mass in the
galaxy can be found using the models of {Li \& Draine (2001,
2002)}\nocite{li01,li02} and \citep{draine06}.  Here, radiation field
strengths are varied via power-law distributions; PAH, silicate and
graphite grains are illuminated and the resulting SEDs can be compared
to the observations. The fit of these models to our observations is
also shown in Figures~\ref{figcap10}--\ref{figcap12}.  The observed
SEDs are well-reproduced by these models, when a small fraction
($\sim$5--30\%, depending on the individual region) of the dust mass
is exposed to radiation fields largely in excess of the local average
(see {Draine \& Li 2006}\nocite{draine06}).  The implied dust masses are
given in Tables~\ref{t3} and \ref{t4}; these masses are smaller than
those derived using the \citet{dale01} and \citet{dale02} models.  The
derived global dust mass is (1.03\,\pom\,0.52)\,$\times$\,10$^{5}$
\msun.

We note that our global flux densities produce SEDs that are in good
agreement with the fluxes measured in \citet{israel96}.  Our use of
sophisticated SED modeling techniques [\citet{dale01} and
\citet{dale02}; {Li \& Draine (2001, 2002)}\nocite{li01,li02} and
{Draine \& Li (2006)}\nocite{draine06}] results in dust masses that are
larger than those derived by \citet{israel96}. These differences are
thus attributable solely to our different (and likely more physical)
modeling approach.

Note that our data may remain insensitive to a potential component of
very cold dust (which may be located away from regions of star
formation).  Previous studies (e.g., {Eales \etal\
1989}\nocite{eales89}, {Devereux \& Young 1990}\nocite{devereux90},
{Kwan \etal\ 1992}\nocite{kwan92}, and {Calzetti \etal\
2000}\nocite{calzetti00}) have shown that observations from the mid-IR
into the sub-millimeter regime are needed to fully quantify the range
of dust temperatures, grain sizes, and opacities, and to provide a
robust estimate of the total dust mass in a galaxy. \citet{popescu02},
for example, noted that $\sim15$-20~K dust may be present in many
nearby galaxies.  Data longward of 160~\mm\ would be needed to
determine whether such dust is present in NGC\,6822.  Some
observational results have also demonstrated that dust with
temperatures as cold as 5-10~K (e.g., {Madden \etal\
2002}\nocite{madden02}; {Galliano \etal\ 2003}\nocite{galliano03},
{2005}\nocite{galliano05}) may be present in dwarf galaxies.  However,
theoretical research (e.g., {Li 2004}\nocite{li04}) and other
observational results \citep[e.g.]{dumke04,bendo06} have suggested
that 5-10~K dust temperatures would require physically implausible
dust grain properties or unrealistically high dust masses.
Nonetheless, if dust cooler than 20~K exists in NGC\,6822, then these
data may afford only a lower limit on the total dust content of the
system.

\subsubsection{The Empirical Dust-to-\HI\ Ratio}
\label{S3.2.3}

The \spitzer\ images of NGC\,6822 can be compared to our sensitive
\HI\ imaging to study the dust-to-gas ratio at $\sim$ 130 pc
resolution.  We define the dust-to-gas ratio (hereafter represented as
\dgr) as the ratio of dust mass (as traced by these \spitzer\
observations) to \HI\ mass; no correction is applied for the
(uncertain) contribution of CO or H$_2$ gas.  At sub-solar
metallicities (recall that NGC\,6822 has a metallicity $\simeq$30\%
\zsun), the molecular phase of the ISM is different from that in more
metal-rich systems.  Theoretical predictions
\citep[e.g.][]{maloney88,bolatto99} and observational evidence
\citep[e.g.,][]{taylor98,leroy05} have shown that CO clouds have a
smaller volume filling factor in metal-poor systems than in spiral
galaxies.  \citet{israel97} uses CO observations of selected regions
in NGC\,6822 to infer a total molecular mass $\sim$10\% of the \HI\
mass (though these numbers are highly uncertain, depending on the
conversion factor used to infer total H$_2$ content from observed CO
intensities). Since a complete census of the molecular material in
NGC\,6822 is difficult, we simply study the relation of dust and
\HI\ gas; any molecular material in NGC\,6822 will lower the derived
\dgr.

Throughout the disk, the mean value of \dgr\ (in apertures of physical
radius $\sim$130 pc, using the dust masses derived from the models of
Li \& Draine) is $\sim$0.004 (with little change when using apertures
with 4 times larger area; see Tables~\ref{t3} and \ref{t4}).
Variations of a factor $\sim$3 are seen among individual regions, and
the errors on the \dgr\ values are also high (arising from the
relatively uncertain dust mass estimates).  These average values are
lower than the average global values found for other
\sings\ galaxies using similar modeling techniques.  For example,
Draine \etal\ (2006, in preparation) present detailed dust mass estimates
for many {\it SINGS} galaxies: for 40 systems with \spitzer\ data but
no submillimeter observations, the average global \dgr\ (excluding dwarf
galaxies and the contribution of molecular hydrogen) is M$_{\rm
dust}$/M $_{\rm HI}$=0.02; i.e. a factor of $\sim$5 higher than the
average local ratios found using the same models in NGC\,6822.

\citet{israel96} finds a global \dgr\ of 1.4\,$\times$\,10$^{-4}$ in
NGC\,6822 using \iras\ data; our derived global value (using the Li \&
Draine models) is $\sim$8\,$\times$\,10$^{-4}$ (see Table~\ref{t3}).
Our {\it Spitzer} data recover very similar global flux densities as
found in the \citet{israel96} study; the differences in dust mass and
\dgr\ are attributed to the use of detailed SED models of the IR
emission.  Note that all of the \dgr\ values derived for the apertures
listed in Table~\ref{t4} are higher than the global value.  This is
expected since the dust is strongly peaked in regions of high \HI\
column and star formation (e.g., see Figures~\ref{figcap5},
\ref{figcap7} and \ref{figcap8}), but the bulk of the extended \HI\
distribution has little associated dust emission (see
Figure~\ref{figcap2}).  Recall, however, that our data remain
insensitive to a potential component of very cold dust (see more
detailed discussion above).

We note by examining the individual panels of Figures~\ref{figcap10},
\ref{figcap11} and \ref{figcap12} that the model fits to the IR SEDs
predict significant power in the PAH bands between $\sim$5 and 20 \mm\
(though see discussion of potential metallicity effects in {Dale
\etal\ 2005}\nocite{dale05}).  Most regions show a rise in the SED
moving from 4.5 \mm\ to 5.8 and 8.0 \mm, part or all of which may be
associated with aromatic band emission in the ISM.  Given the
metallicity of NGC\,6822 (Z $\simeq$ 30\% Z$_{\odot}$) and the
empirical dependence of PAH emission on metallicity
\cite{engelbracht05}, it will be fruitful to compare the IRAC images
with IRS spectra of various regions in NGC\,6822. Initial results of
\sings\ extra-nuclear IRS spectra of the major star formation
complexes (e.g., Hubble I/III, V, X) show that PAH emission is
prominent throughout the galaxy.  This will
allow a proper spatial and spectral decomposition of PAH emission in
the metal-poor ISM at unprecedented resolution.

\subsection{The Radio - Far-IR Correlation}
\label{S3.3}

In normal star-forming galaxies, there is a remarkably tight
correlation between the total IR luminosity and the strength of radio
continuum emission (see, e.g., {de~Jong \etal\ 1985}\nocite{dejong85},
{Helou \etal\ 1985}\nocite{helou85}, {Condon 1992}\nocite{condon92},
and references therein).  Given the differing physical mechanisms
giving rise to far-IR and radio emission, the tightness of the
relation (scatter of less than 0.3 dex over multiple orders of
magnitude in luminosity) is especially remarkable.  \citet{bell03a}
shows that the relation holds for dwarf galaxies, which are generally
deficient both in IR and radio emission with respect to massive
galaxies, when the emission is normalized by the star formation rate.
However, these systems require a complex interplay of star formation,
dust content and magnetic fields to produce the apparently linear
relation in this regime.

Figure~\ref{figcap3}(h) shows that only sources 2, 3, 4, 7, 8, and 14
are detected at high S/N in our radio continuum imaging. These are the
six highest-surface brightness \halpha\ regions in the galaxy.  We
tentatively detect lower surface brightness regions throughout the
disk, but the presence of strong background sources precludes us from
analyzing them with confidence.  Flux densities for these regions are
quoted as upper limits in Table~\ref{t2}.  Assuming that the
monochromatic or TIR fluxes and the radio continuum emission are
related via constant values [q $\propto$ log(S$_{\rm IR}$/S$_{\rm
RC}$); see {Bell (2003)}\nocite{bell03a} and {Murphy \etal\
(2006a)}\nocite{murphy06a} for details], we calculate two values of
``q''.  The first is a monochromatic comparison of the IR and radio
continuum flux densities:
\begin{equation}
q_{\lambda\ (\rm \mu m)} = log\frac{{\it f}_{\nu}(\lambda) (Jy)}{{\it f}_{\nu}(20 cm) (Jy)}
\label{eq}
\end{equation}
\noindent The second compares a measure of the total infrared flux 
with the radio continuum:
\begin{equation}
q_{\rm TIR} =log\frac{TIR}{(3.75E12\,\,W\,\,m^2)} - log\frac{S_{1.4 GHz}}{(W\,\,m^2\,\,Hz)}
\end{equation}
We calculate ``q'' (see Table~\ref{t3}) in individual regions, as well
as throughout the total galaxy.  The resulting values are in excellent
agreement with those found both locally and globally in larger samples
of galaxies.  For example, \citet{bell03a} finds a median global
q$_{\rm TIR} =$ 2.64\pom0.02 in a sample of more than 200 galaxies.
\citet{murphy06a} explores the radio-IR correlation on a spatially
resolved basis (sampling physical regions of $\sim$300--750 pc) using
\sings\ data on four nearby spiral galaxies, finding ranges of
0.85\,$\lsim$\,q$_{\rm 24}$\,$\lsim$\,1.10 and 1.95\,$\lsim$\,q$_{\rm
70}$\,$\lsim$\,2.25 (where q$_{\rm 24}$ and q$_{\rm 70}$ are the
monochromatic ``q'' parameters derived using equation~\ref{eq}).  In
NGC\,6822, there is slight evidence that regions with a high
\halpha/TIR ratio show lower ``q'' values; there is no evidence for a
trend of ``q'' with \dgr\ or with TIR flux.

The agreement of the local and global values with the canonical
radio-IR relations derived for spiral galaxies suggests that both the
radio and IR are significantly depleted in NGC\,6822, conspiring to
produce the observed ``q'' parameters.  Different physical mechanisms
appear to be at work in the ISM of this dwarf galaxy compared to those
operating in the ISM of more massive spirals.  In NGC\,6822, the bulk
of the radio emission appears to be thermal: in \S~\ref{S3.1.1}, we
showed that most regions have thermal fractions $>$90\% (the strongest
nonthermal sources are regions 7 and 8, with thermal fractions of
$\sim$50\% and 70\%, respectively).  In contrast, nonthermal processes
produce the bulk of radio continuum emission in massive spiral
galaxies [see {Murphy \etal\ (2006a)}\nocite{murphy06a} and {Murphy
\etal\ (2006b)}\nocite{murphy06b} for more detailed discussions of the
radio-FIR correlation in \sings\ galaxies].  Evidently, in NGC\,6822
the IR emission per unit star formation is reduced by about the ratio
of free-free/nonthermal emission in normal spirals, conspiring to
produce the observed ``q'' values.

The relative ratios of radio and IR emission strengths in NGC\,6822
are in marked contrast to the strong deviations from the canonical
radio-IR relation found for very active low-metallicity galaxies in
the \sings\ sample.  \citet{cannon05,cannon06} show that the radio-FIR
correlation breaks down in very strongly star forming regions.  The
``supergiant shell'' region of IC\,2574 has induced strong variations
of the ``q'' parameter (over an order of magnitude); the dwarf
starburst galaxy NGC\,1705 shows strong FIR emission but is a radio
non-detection.  Other types of galaxies can also be extremely
radio-deficient, though likely for different reasons than those seen
in NGC\,6822 [see, e.g., the {Roussel \etal\ (2006)}\nocite{roussel06}
study of the nascent starburst in the lenticular galaxy NGC\,1377].

These results suggest that star formation intensity has a dramatic
effect both on the appearance of radio continuum emission, and on the
derived correlation between FIR emission and radio continuum
luminosity, in galaxies with small potential wells (for comparison,
note that more massive starburst galaxies in the {Bell
2003}\nocite{bell03a} sample show little deviation from the derived
radio-TIR relation).  The intensity of star formation relative to the
presence of a stable large-scale disk is important, since the ratio of
activity to disk density and gravity will govern the amount of
disruption and blow-out that takes place (thus, the irregular nature
of NGC\,6822 is likely also important in determining the ratio of IR
to radio emission).  The relatively uniform values of ``q'' found in
larger spiral galaxies and quiescent dwarfs argues that disk
disruption by star formation is minimal; the strongly varying values
of ``q'' in starbursting dwarfs suggests that the small disks are
being overwhelmed by the star formation activity.

\section{Discussion and Conclusions}
\label{S4}

We have presented IR observations of the Local Group dwarf irregular
galaxy NGC\,6822 obtained with {\it Spitzer} as part of \sings.  The
galaxy is highly resolved at all imaging bands; the resolution limit
is driven by the FWHM of the 160 \mm\ PSF.  At $\sim$130 pc physical
resolution, we study the nature of FIR emission and compare to
observations in the optical and radio. These sensitive data reveal a
wealth of structure within the galaxy.  Our study confirms some
previous results, while also offering new insights into the nature of
infrared emission in the metal-poor ISM.  We discuss these topics in
more detail below.

The total monochromatic FIR flux densities are found to be S$_{\rm 24}
=$ 2.51\,$\pm$\,0.50 Jy; S$_{\rm 70} =$ 53.2\,$\pm$\,15 Jy; S$_{\rm
160} =$ 136.2\,$\pm$\,40 Jy.  SED model fits to these data allow
comparison with previous {\it IRAS} flux measurements; our {\it
Spitzer} flux densities are in excellent agreement with the
measurements in \citet{israel96}.  Using the relations presented in
\citet{dale02}, these global flux densities correspond to a total IR
flux of $\sim$ 5.7\,$\times$\,10$^{-12}$ W\,m$^{-2}$ ($\sim$
4.3\,$\times$\,10$^7$ L$_{\odot}$).

IR emission at 24, 70 and 160 \mm\ is found only within the central,
high-\HI\ column density ($\gsim$ 10$^{21}$ cm$^{-2}$) region of
NGC\,6822. The total \HI\ distribution is much more extended than the
high-surface brightness stellar or dust emission components.  The 24
\mm\ surface brightness profile closely traces that of \halpha,
delineating regions of ongoing star formation.  The profiles of 70 and
160 \mm\ emission also trace local \HI\ column density maxima.

Comparing (monochromatic and total) FIR emission with a sensitive
\halpha\ image reveals a strong morphological correlation between IR
and \halpha\ surface brightness; here we are directly observing the
re-processing of photons by dust.  Roughly 50\% of the TIR flux is
attributable to discrete FIR sources, all of which have at least a
low-level \halpha\ counterpart; the remaining FIR emission arises from
a ``diffuse'' component of the ISM (slightly stronger at 70 and 160
\mm\ than at 24 \mm; see detailed discussion in \S~\ref{S3.1.1}) that
is apparently not associated with massive star formation.
\citet{gallagher91} and \citet{israel96} find similar results from
{\it IRAS} imaging of NGC\,6822; we interpret this as the
re-processing of non-ionizing photons in the ISM or as the escape of
radiation from the star formation regions.

We find variations in the relative ratios of nebular and dust
continuum emission. While differential extinction can cause some of
the observed fluctuations, it cannot account for the full breadth of
properties seen in the galaxy.  Using \sings\ observations of the
starbursting dwarf galaxies IC\,2574 and NGC\,1705,
\citet{cannon05,cannon06} found similar variations and attributed them
to the direct impact of recent star formation on the local ISM (i.e,
production and re-processing of photons).  The results presented here
extend this trend to more typical, quiescent dwarfs.

Evidence is mounting from spatially resolved {\it Spitzer}
observations of dwarf galaxies that these systems follow a different
trend of 24 \mm\ vs. \halpha\ luminosity than what is seen in more
metal-rich systems (see, e.g., {Calzetti \etal\
2005}\nocite{calzetti05}).  This result is in agreement with previous
observations of dwarf galaxies \citep[e.g.,][]{hunter89}.  Two
different scenarios can account for this observational trend. The
first one, physically simple and more commonly used, is related to the
low dust content of low-metallicity galaxies. Here, low-metallicity
galaxies simply do not have enough metals to form dust in the same
fraction as more metal rich environments.  The second scenario posits
that dwarfs harbor massive components of cold dust (T $\sim$ 5--10 K)
that is simply not heated by ongoing star formation. Such dust would
peak past 200 \mm\ and remain undetected in the present \spitzer\
observations, contributing to part of the observed discrepancy;
observational evidence exists both for and against such a component in
nearby galaxies (see detailed discussion in \S~\ref{S3.2.2}).
Clearly, sensitive observations of a large sample of dwarf galaxies
between 200 and 1000 \mm\ are needed to address this issue.

As mentioned previously, our global flux densities for NGC\,6822 are
in good agreement with those from \citet{israel96}.  We use
sophisticated SED modeling techniques [\citet{dale01} and
\citet{dale02}; {Li \& Draine (2001, 2002)}\nocite{li01,li02} and
{Draine \& Li (2006)}\nocite{draine06}] to derive dust masses that are
higher than those derived by \citet{israel96}.  The differences in
mass are thus attributable solely to our more physical modeling
approach, which takes into account the variety of dust species and
heating levels in each region within the disk (see detailed discussion
in \S~\ref{S3.2.2}).  For a given TIR luminosity, our models recover a
higher dust mass than simple blackbody fits (see detailed discussion
in \S~\ref{S3.2.2}).  The treatment of multiple grain populations
likely means that the increased dust masses are a better
representation of the mass radiating between 10 and 160 \mm. Using
these techniques, we derive a total dust mass of $\sim$10$^5$ \msun\
(this value is derived using the Li \& Draine models).
 
Comparing to our high-resolution \HI\ imaging, we can explore the mass
ratio of warm dust to \HI\ gas (\dgr) in the ISM.  Given the low
nebular abundance of the system (Z $\simeq$ 30\% \zsun), we can thus
provide constraints on a potential trend of \dgr\ with abundance.
Metal-poor dwarf galaxies might be expected to have a lower \dgr\
(i.e., higher gas mass fraction) than more metal-rich galaxies such as
the Milky Way \citep[e.g., ][]{thronson86}.  \citet{lisenfeld98} have
argued that dwarf irregular galaxies have a clear relation between
these two quantities in a global sense (using dust masses derived from
single-temperature blackbody fits).

Here, we derive a global \dgr\ $\simeq$ 8\,$\times$\,10$^{-4}$.  This
is $\sim$25 times lower than the average global values found in the
new \sings\ study of Draine \etal\ (2006, in prep.), wherein dust
masses are calculated for a large sample of galaxies (excluding
dwarfs).  On local scales ($\sim$130 pc), the dust mass per unit \HI\
mass is higher (by a factor of $\sim$5), though still lower than the
global values found by Draine et\,al.

We propose a simple model for the relative ratios of dust mass to \HI\
mass in the ISM.  As noted above, dwarf galaxies may contain large
reservoirs of cold dust (T $\sim$ 5--10 K) that are undetectable with
\spitzer\ observations \citep[e.g.][]{madden02,galliano05}.  Note,
however, that the dilution of the radiation field required for this
dust to remain at such low temperatures may require it to be located
at large distances from the star formation regions and thus
essentially decoupled from the active ISM (note that other mechanisms
may also be important in dwarfs, including dust destruction by
supernova shocks; see, e.g., {Bot \etal\ 2004}\nocite{bot04}).
Superposed on this putative cool component are regions radiating
between 10 and 160 \mm; here, photons are being absorbed and
re-radiated as IR photons.  The low dust contents of metal-poor
galaxies plays an important role in the microphysics of this process.
Galaxies with widespread, elevated star formation will heat more of
their total dust component than more quiescent systems.  When more
sensitive observations are possible between 200 and 1000 \mm, we will
be able to quantify the relative masses of these components in
low-mass dwarf galaxies.

Finally, it is interesting that NGC\,6822 seems to fall near the
canonical measures of the radio-FIR relation
\citep[e.g.,][]{bell03a,murphy06a}.  20\,cm radio continuum imaging
reveals six high surface brightness sources, each of which appears to
be of thermal origin. The monochromatic and total IR ``q'' parameters
are similar to those found for more massive galaxies.  This agreement
is interesting, given that nonthermal emission dominates the radio
continua of spiral disks.  The simplest explanation is that depleted
radio and far-IR emission strengths conspire to produce the observed
values.  These findings in NGC\,6822 are in marked contrast to recent
results for very active star-forming dwarf galaxies, where significant
deviations from the canonical radio-FIR relation are found
\citep{cannon05,cannon06}.  Taken together, these data suggest that
star formation intensity is one of the important parameters that
governs the observed radio-FIR relation in dwarf galaxies.

\acknowledgements

Support for this work, part of the {\it Spitzer Space Telescope}
Legacy Science Program ``The Spitzer Nearby Galaxies Survey,'' was
provided by NASA through contract 1224769 issued by JPL/Caltech under
NASA contract 1407. The authors thank the anonymous referee for a
constructive review, and Henry Lee for for providing optical
spectroscopy results prior to publication.  This research has made use
of the NASA/IPAC Extragalactic Database (NED) which is operated by the
Jet Propulsion Laboratory, California Institute of Technology, under
contract with the National Aeronautics and Space Administration, and
NASA's Astrophysics Data System.

\end{document}